\documentclass[11pt]{article}

\usepackage{amsmath,amssymb}
\usepackage{amsfonts}
\let\sect=\section
\def\section{\newpage\sect}

\def\text#1{\mbox{\rm #1\ }}

\def\tr{{\rm Tr}}
\def\diag{{\it Diag}}
\def\ie{{\rm i.e.,\/}\ }
\def\etc{{\rm etc.\/}\ }

\def\id{\mbox{\it id\,}}
\def\one{\mbox{\rm 1}\hskip-2.8pt \mbox{\rm l}}
\newcommand{\Mlo}[1]{\ensuremath{(M_{#1}(\Lambda^{2}))_0}}
\newcommand{\ZZ}{\mathbb{Z}}
\newcommand{\RR}{\mathbb{R}}
\newcommand{\CC}{\mathbb{C}}
\title{Hopf stars, twisted Hopf stars and scalar products 
       on quantum spaces
   \vspace{0.8cm}
}

\author{R. Coquereaux${}^1$\thanks{~Email: coque@cpt.univ-mrs.fr}$\;$,
        A. O. Garc\'{\i}a${}^1$\thanks{~Email: garcia@cpt.univ-mrs.fr}$\;$,
        R. Trinchero${}^2$\thanks{~Email: trincher@cab.cnea.gov.ar} \\
\\
${}^1$ {\it Centre de Physique Th\'eorique - CNRS} \\
       {\it Campus de Luminy - Case 907}           \\
       {\it F-13288 Marseille - France}            \\
\\
${}^2$ {\it Instituto Balseiro and Centro At\'omico Bariloche}         \\
       {\it CC 439 - CP 8400 - Bariloche - R\'{\i}o Negro - Argentina} \\
\\
}

\date{}

\begin{document}

\begin{titlepage}
\thispagestyle{empty}

\maketitle

\vfill

\abstract{
The properties of Hopf star operations and twisted Hopf stars operations 
on quantum groups are discussed in relation with the theory of
representations (star representations). Invariant Hermitian sesquilinear 
forms (scalar products) on modules or module-algebras are then defined 
and analyzed. Particular attention is paid to scalar products that can 
be associated with the Killing form (when it exists) or with the left 
(or right) invariant integrals on the quantum group.

Our results are systematically illustrated in the case of a family
of non semi-simple and finite dimensional quantum groups that are 
obtained as Hopf quotients of the quantum enveloping algebra 
$U_q(sl(2,\CC))$, $q$ being an $N$-th root of unity. Many explicit 
results concerning the case $N=3$ are given.

We also mention several physical motivations for the present work: 
conformal field theory, spin chains, integrable models, generalized 
Yang-Mills theory with quantum group action and the search for finite 
quantum groups symmetries in particle physics.
}

\vspace{0.8 cm}

\noindent PACS: 02.90.+p, 11.30.-j \\
\noindent MSC: 16W30, 81R50 \\
\noindent Keywords: quantum groups, Hopf algebras, star structures,
                    scalar products, non commutative geometry.

\vspace{0.5cm}

\noindent Anonymous ftp or gopher: cpt.univ-mrs.fr

\vspace{0.3 cm}

\noindent {\tt math-ph/9904037}\\
\noindent CPT-99/P.3808 \\
\noindent %% IT-CNEA-CAB/*****

\vspace*{0.3 cm}

\end{titlepage}

%%%%%%%%%%%%%%%%%%%%%%%%%%%%%%%%%%%%%%%%%%%%%%%%%%%%%%%%%%%%%%%%%%%%%%%%%% 
%%%%%%%%%%%%%%%%%%%%%%%%%%%%%%%%%%%%%%%%%%%%%%%%%%%%%%%%%%%%%%%%%%%%%%%%%% 

\section{Introduction}

The purpose of the present paper is to study the concepts of Hopf star 
operations and twisted Hopf star operations in the theory of quantum 
groups. This study is motivated by a number of physical considerations 
that we shall discuss in this introduction.

First of all, one should remember that the notion of quantum group 
(Hopf algebra) does not make use of a star operation ---roughly 
speaking, the notion of complex conjugate---; chosing one comes only 
at a later stage. Such an operation is an anti-multiplicative and 
anti-linear involution which could be quite arbitrary when the 
(associative) algebra under consideration is not a Hopf algebra.
 
However, the existence of a coproduct allows one to distinguish two 
particular kinds of star operations. The problem is to relate the star 
operations that one can define on the algebra $H$ and on its tensor 
square $H\otimes H$, since we have a very special embedding of the 
first algebra into the latter one given by the coproduct. If 
$\Delta a = a_1\otimes a_2$, it may be that the chosen star is such 
that $\Delta (a^*) = a_1^*\otimes a_2^*$ (a Hopf star operation) but 
it also could happen that $\Delta (a^*) = a_2^*\otimes a_1^*$ (a 
twisted Hopf star operation).

Actually, one could define also ``partially twisted stars'' which in 
a sense continuously interpolate between a Hopf and a twisted star 
(see \cite{Buffenoir-Roche}), but these involve additional data, an 
element $f \in H \otimes H$.

In the case of Lie groups or Lie algebras, star operations are used to 
define real forms. However, for Hopf algebras the notion of ``real 
form'' is slightly more subtle (we shall say more about it later), 
but it is a priori clear that the notions of complex conjugate and of 
star representations should be discussed as soon as one wants to endow 
a representation space with some sort of scalar product.

A general discussion of star versus twisted star operations seems to be 
lacking in the literature: mathematical books on quantum groups (for 
instance \cite{Chari-Pressley} or \cite{Klimyk}) only discuss (genuine) 
Hopf star operations, the same being true for all research papers 
studying $C^*$-algebra aspects of ``matrix quantum groups'' (in the 
sense of Woronowicz \cite{Woronowicz}). In the physics literature, most 
papers dealing with applications of quantum groups to integrable models, 
spin chains, or conformal field theory, usually do not choose any 
particular star operation at all on the quantum group of interest. But 
sometimes they do, and it turns out that the chosen star is often a 
twisted star ---although usually the authors do not acknowledge the 
fact that it is so\footnote{
With the notable exception of papers by G. Mack and V. Schomerus
\cite{Mack-Schomerus-1, Mack-Schomerus-2}.
}, 
and this state of affairs creates some confusion. Quantum groups have 
also been discussed in relation with the possibility of $q$-deforming 
the Lorenz group, and here again, the two possibilities (twisted 
versus non twisted) appear in the physical literature: from one side 
we have the papers \cite{Wess-Zumino-Ogievetsky} or 
\cite{Podles-Woronowicz}, whereas from the other we have the 
articles \cite{Lukierski-et-al}.

Another motivation for our work comes from the possibility, as 
advocated by A. Connes in \cite{Connes}, that reduced quantum groups 
like $U_q(sl(2,\CC))$ at a cubic root of unity could have some 
essential role to play in the formulation of fundamental interactions 
(Standard Model). This suggestion is based upon the following two 
observations: first of all, when $q$ is chosen to be a cubic root of 
unity the algebra of ``functions'' $Fun(SL_q(2,\CC))$ is a Hopf-Galois 
extension of $Fun(SL(2,\CC))$ ---the algebra of complex valued functions 
on the Lorenz group---, the fiber being a finite dimensional quantum 
group $\mathcal{F}$ whose Hopf dual is a finite dimensional Hopf algebra 
quotient $U^{res}_q(sl(2,\CC))$ (that we shall call $\mathcal{H}$) of 
the quantum enveloping algebra $U_q(sl(2,\CC))$. Next, for this $q$ the 
semi-simple part of $\mathcal H$ turns out to be isomorphic with the 
algebra $M(3,\CC)\oplus M(2,\CC) \oplus \CC\,$. It is then tempting to 
use the tools of non commutative geometry to build a physical model 
that would recover the usual Standard Model ---maybe a generalization 
of it---, incorporating some action of an hitherto unnoticed finite 
quantum group of symmetries. The existence of a non trivial coproduct 
mixing the different components of $\mathcal{H}$ and the nature of the 
representations of this non semi-simple Hopf algebra make it quite hard 
to recover the usual model of strong and electroweak interactions; this 
has not been achieved yet. In any case, it is clearly of interest to 
analyze in detail the structure of the representation theory of this 
Hopf algebra, and to pay particular attention to the different kinds 
of ``reality'' structures that one can find for these representations.
For these reasons, and although we decided to write quite a general 
paper, most explicitly discussed examples will involve the case of the 
finite dimensional algebra $\mathcal{H} = U^{res}_q(sl(2,\CC))$ at a 
cubic root of unity.

Another motivation for studying the reality structures and the type of 
scalar products in star representations of quantum groups comes from 
our previous work \cite{CoGaTr-l, CoGaTr-e}. Here a new kind of gauge 
fields was obtained: starting from the observation that the reduced 
quantum plane (identified with the algebra of $N\times N$ complex 
matrices) is a module-algebra for the finite dimensional quantum group 
$\mathcal{H}$, when $q^N=1$, we built a differential algebra over it by 
taking an appropriate quotient of the Wess-Zumino differential algebra 
over the ---infinite dimensional--- quantum plane; generalized 
differential forms are then obtained by making the tensor product of 
the De Rham complex of forms over an arbitrary space-time manifold times 
the previous Wess-Zumino reduced differential complex; generalized gauge 
fields (and curvatures, \etc) are finally constructed by standard non 
commutative geometrical techniques. Clearly, the wish to construct a 
lagrangian model involving the representations of a quantum group (that 
knows how to act on such generalized gauge fields) requires the study of 
star (or twisted star) operations on the corresponding modules.

Finally, the last motivation comes from spin chains, integrable models 
and conformal theories. The $q$-parameter appearing in many conformal 
field theory models and integrable models is a primitive root of unity. 
Such values, as a rule {\sl exclude} the choice of a Hopf star operation 
leading to a compact quantum group like $SU_q(2)$, for instance. For this 
reason star operations used in articles like \cite{Pasquier-Saleur} 
---where the role of quantum groups is discussed in the context of spin 
chains, like $SU_q(2)$ in the $XXZ$ model--- are not true Hopf star 
operations; we shall return to this discussion in 
Section~\ref{s:discussion}.

%%%%%%%%%%%%%%%%%%%%%%%%%%%%%%%%%%%%%%%%%%%%%%%%%%%%%%%%%%%%%%%%%%%%%%%%%% 

\bigskip

The structure of our paper is the following:
\smallskip

In Section~\ref{s:stars} we gather information on stars operations: Hopf 
and twisted Hopf stars, compatible stars on modules and module-algebras, 
behaviour under tensor product of representations, \etc

In Section~\ref{s:invariant_scalar_products} we discuss scalar products 
in representation spaces, its quantum invariance and associated star 
representations. As everywhere else in this paper, we first discuss 
all the general notions and then exemplify by taking the finite 
dimensional quantum group $\mathcal{H} = U^{res}_q(sl(2,\CC))$ for 
$q$ a primitive odd root of unity (most of the time we take $N=3$). 
The characteristics of the invariant scalar products on the irreducible 
and the projective indecomposable representations of $\mathcal{H}$ are 
studied in detail, both in the case where a genuine or a twisted Hopf 
star is chosen on $\mathcal{H}$. The same analysis is carried out for 
the module-algebra $M(N,\CC)$.

In Section~\ref{s:scalar_products_on_left_rr} we examine more particularly 
the (left) regular representation of a Hopf algebra $H$ and exhibit two 
distinguished invariant scalar products. The first one is defined in terms 
of the Killing form. The other is built using the left (or right) invariant 
integral on the algebra $H$. We then analyze in detail these scalar 
products for the case of $\mathcal{H}$. As we shall see, it happens that 
for many properties Hopf stars behave usually much better than twisted Hopf 
stars.

\ref{a:structure_of_H} summarizes what is needed for this paper from 
the structure and representation theory of the finite dimensional Hopf 
algebras $\mathcal{H} = U^{res}_q(sl(2,\CC))$ when $q$ is an odd 
primitive root of unity, in particular the structure of the projective 
indecomposable modules and of the corresponding irreducibles.

\ref{a:Killing_form} recalls a few properties concerning the adjoint 
representation of quantum groups, together with the notions of quantum 
trace and quantum Killing form.

\ref{a:algebras_related_to_H} gives a few explicit results concerning a 
``double cover'' of the finite dimensional Hopf algebra $\mathcal{H}$.

\subsection*{About notations}

$F$ will generically denote a complex Hopf algebra, for example the 
algebra of ``functions'' on a quantum group. $H$ will be its dual (also 
a Hopf algebra), so that it can be thought of as a non commutative 
generalization of the group-algebra of a finite group or as the 
non-commutative analogue of the enveloping algebra of a Lie algebra. 
As already mentioned, the particular examples where $H$ is chosen to 
be one of the finite dimensional quotients of $U_q(sl(2,\CC))$ will be 
called ${\mathcal H}$. $V$ will denote a representation space for $H$ 
(and we shall have to specify if it is a left or a right module), and 
will therefore also be a (left or right) co-representation space of the 
Hopf algebra $F$. Finally, $M$ will denote a module-algebra for $H$ 
(\ie a comodule-algebra for $F$).

%%%%%%%%%%%%%%%%%%%%%%%%%%%%%%%%%%%%%%%%%%%%%%%%%%%%%%%%%%%%%%%%%%%%%%%%%% 
%%%%%%%%%%%%%%%%%%%%%%%%%%%%%%%%%%%%%%%%%%%%%%%%%%%%%%%%%%%%%%%%%%%%%%%%%% 
\section{Stars}
\label{s:stars}

\subsection{Hopf stars}

Remember that a star on an algebra is an involutive antilinear 
antiautomorphism, \ie
$$
\begin{array}{l}
\left( x^*\right)^* = x \\
(\lambda \, x)^* = \bar{\lambda} x^* \;, \quad \quad \lambda \in \CC 
\\
(xy)^* = y^* x^*
\end{array}
$$

\noindent
Now let the algebra on which $*$ acts be a complex Hopf algebra 
$H(m,\Delta,\eta,\epsilon,S)$. In this case one requires the star to 
satisfy two extra compatibility conditions \cite{Chari-Pressley} with 
the Hopf operations:
\begin{eqnarray}
   \Delta \, * &=& *_{_{\otimes}} \, \Delta 
      \label{normal_Hopf_star} \\
   \epsilon \,* &=& *_{_{\CC}}\,\epsilon \ . \nonumber
\end{eqnarray}

\noindent
However the $*$'s on the right hand side are operators on different 
spaces and are yet to be defined. $*_{_{\CC}}$ should be a star on 
$\CC$, and therefore is just complex conjugation. The operation 
$*_{_{\otimes}}$ should be an involution on $H \otimes H$, the 
standard choice is
$$
   *_{_{\otimes}} = * \otimes * \ .
$$

\noindent
A star satisfying (\ref{normal_Hopf_star}) with the standard choice 
of $*_{_{\otimes}}$ is called a Hopf star, and in such a case $H$ is 
called a Hopf star algebra.

Actually one could also make the choice
$*_{_{\otimes}} = \tau(* \otimes *)$,
where $\tau$ is the tensorial flip (twisting); however, making such a 
choice and imposing (\ref{normal_Hopf_star}) amounts to make the 
standard choice for $*_{_{\otimes}}$ and rewrite 
(\ref{normal_Hopf_star}) as
$$
   \Delta \, * = *_{_{\otimes}} \, \Delta^{op} \ , 
$$
where $\Delta^{op} \doteq \tau \circ \Delta$ is the opposite coproduct. 
We will call this second type of operation a {\it twisted} Hopf star, 
or even a twisted star. In this paper, therefore, we shall always make 
the standard choice for $*_{_{\otimes}}$. In this section we will 
analyze Hopf star algebras, leaving the study of the twisted case to 
Section~\ref{s:twisted_star}. 

Remark that there is no need to impose a relation between the star 
and the antipode (which is a {\sl linear} antiautomorphism) because 
this one arises automatically. In fact, it is easy to see that 
\begin{eqnarray}
   S \,* \, S \, * &=& \id \ .
\end{eqnarray}

\noindent
This is so because $* \,S^{-1}\, * = (*\,S\,*)^{-1}$ satisfies all 
the conditions for the antipode, which is unique. We should therefore 
remember that for Hopf star algebras
\begin{eqnarray*}
S \,* &=& * \, S^{-1} \ .
\end{eqnarray*}
Notice that in general $S$ has no reason to be equal to $S^{-1}$ 
(imposing such a property would exclude all the Drinfeld-Jimbo 
deformations!).

Given a Hopf algebra $H$, one can consider its dual\footnote{
The examples that we shall consider in this paper are finite 
dimensional (and non semisimple) Hopf algebras, therefore it 
will be possible to identify canonically a given Hopf algebra 
with its bidual.
}
Hopf algebra $F=H^\star$ with operations such that 
\begin{eqnarray}
   \langle \Delta f, h \otimes h' \rangle &=& \langle f\,, hh'\rangle 
      \qquad\qquad \forall \:h,h'\in H \nonumber \\
   \langle ff', h \rangle &=& \langle f\otimes f' \,, \Delta h\rangle \\
   \langle Sf, h \rangle &=& \langle f\,, Sh\rangle \nonumber \\
   \epsilon(f) &=& \langle f\,, 1\rangle \nonumber \\
   \langle \one_F , h \rangle &=& \epsilon(h) \ , \nonumber
\end{eqnarray}

\noindent
where $\langle \: , \: \rangle: F \otimes H \rightarrow \CC$ is the 
bilinear evaluation pairing. When $H$ is a Hopf star algebra, one may 
also define a dual star on $F$. By dual star we mean a star on $F$ 
which is also a Hopf star. It is easy to verify that the following 
formula defines such an operation:

\begin{equation}
\langle f^{\,*},h\rangle =\overline{\langle f\,,(Sh)^{*}\rangle} \ , 
\qquad \forall \:h\in H \ .
\end{equation}

In what follows $F$ will be thought of as the space of functions on a 
quantum group, and its dual $H$ as the quantum group analog of the 
corresponding group algebra (or the enveloping algebra). 

Remark that another standard accepted terminology for denoting the 
star structure of a (untwisted) Hopf star algebra is ``real form on 
a Hopf algebra''. This name does not imply, and we do not construct 
here, any {\it real} Hopf subalgebra of $H$, in the sense of being 
an algebra over the field $\RR$ of real numbers (see 
\cite{Buffenoir-Roche} for a discussion of this point). Let $T$ 
be a {\sl linear} involutive Hopf algebra {\sl anti\/}automorphism
(we call it $T$ for transposition like in \cite{Rosso})
of a complex Hopf star algebra $H$, and consider the subspace
$H_{\RR} \doteq \{h \in H / h^{*} = T(h)\}$. Suppose moreover 
that $H = H_{\RR}\oplus i H_{\RR}$, that $T * = * T$ and 
that $H_{\RR}$ is invariant by the coproduct $\Delta$ (\ie 
$\Delta H_{\RR}\subset H_{\RR} \otimes H_{\RR}$), then $H_{\RR}$ is 
a {\it real} Hopf algebra associated with the star $*$ and the 
involution $T$. Notice that $c \doteq T \, *$ is an {\sl anti\/}linear 
involutive automorphism\footnote{
In the case of our favorite example $\mathcal{H}$, such an operator $c$
can be defined \cite{Kastler}, on the generators, by setting 
$c(X_+) = -q KX_-$, $c(X_-) = -q K^{-1}X+$, $c(K) = K$.
}
and that $H_{\RR}$ is the set of elements of 
$H$ that are invariant under the conjugation $c$. Notice also that if 
$h \in H_{\RR}\:$, then $ih$, as defined in $H$, cannot belong to 
$H_{\RR}$ since $(ih)^{*} = - T(ih)$. When $H$ is ``classical'' (the 
enveloping algebra of some complex Lie algebra), such a $H_{\RR}$ is 
the enveloping algebra of a real Lie algebra. Moreover, in this case 
one takes $T = S$ (since $S^{2}=\id$), so in $H_\RR$ we have 
$x^{*}=T(x) = S(x) = x^{-1}$ for group-like elements and 
$x^{*}=T(x) = S(x) = -x$ for primitive elements. 

%%%%%%%%%%%%%%%%%%%%%%%%%%%%%%%%%%%%%%%%%%%%%%%%%%%%%%%%%%%%%%%%%%%%%%%%%% 
\subsubsection{Selfconjugate representations and compatible stars on 
               modules}

Suppose now that we are given a star $*_H$ on the Hopf algebra $H$, 
and a representation on a vector space $V$. We may have to face 
possible situations. 

The first possibility is that we may want to define a star $*_V$ on 
$V$ and decide to constrain it by imposing some sort of compatibility 
with the star $*_H$ on the quantum group. The second possibility is 
to suppose that we already start with a star $*_V$ on $V$ (a priori 
given);
in such a case\footnote{And thinking now only in the Hopf star case, 
as it is the only one where this notion makes sense.} one can define 
on the same vector space a new representation called the conjugate 
representation. It {\em may} happen that both actions ---the original 
one and its conjugate--- are equivalent. In this last case the 
representation is therefore called {\it self-conjugated}.

Actually, the compatibility condition (see below) between the stars 
in the first scenario is just a particular case of the second option, 
as we define $*_V$ to be such that the representation precisely 
coincides with its conjugate.

Going back to our first problem,
suppose now that we want to define a star $*_V$ on $V$, which is a 
representation space for the quantum group $H$ and a corepresentation 
space for its dual $F$ (\ie $V$ is a right $F$-comodule). Call the 
coaction $\delta_R : V\mapsto V\otimes F$. 

For a Hopf star $*_F$ on $F$ it can be checked that the operation 
$\delta^\prime \doteq (* \otimes *) \, \delta_R \, * \, : V 
\longrightarrow V \otimes F$ is again a right coaction on $V$. 
Therefore it is natural to impose $\delta^\prime = \delta_R$ as the 
compatibility condition between the stars $*_F$ and $*_V$. With a 
slight abuse of notation we can even write 

\begin{equation}
\delta_R (z^*) = \left( \delta_R z \right)^* \ , \qquad z \in V \ , 
\label{Hopf_star_on_comodule}
\end{equation}

\noindent
where the conjugation on the right hand side is the natural star 
structure on $V \otimes F$. In this case we may say that the star is 
covariant.

$V$ being a (right) $F$-comodule, it is also a (left) $H$-module. We 
have indeed an action $\triangleright : H \otimes V \mapsto V$ given 
by
$$
h \triangleright z =
(id \otimes \langle h, \cdot \rangle) \delta_{R}(z) \ . $$
Pairing equation~(\ref{Hopf_star_on_comodule}) with an element $h \in 
H$, and using the duality of real structures we get the equation 

\begin{equation}
h \triangleright z^* = \left[ (Sh)^* \triangleright z \right]^* \ , 
\qquad z \in V \ .
\label{Hopf_star_on_module}
\end{equation}

\noindent
Assuming nondegeneracy of the duality pairing both expressions are 
completely equivalent, and imply some restrictions on $*_V$ given 
$*_F$ or $*_H$.

The action $h \,\triangleright$ of $h$ on $V$ is implemented thanks 
to an endomorphism $\rho[h]$ of this vector space, so one may also 
write $h \,\triangleright \doteq \rho[h]$. Using this notation, 
equation (\ref{Hopf_star_on_module}) can also be written $$
\rho[h](z) = \bar\rho[h](z) \ ,
$$
where $\bar\rho$ denotes the conjugate representation $$
\bar\rho[h](z) = \left[ \rho[(Sh)^*] (z^{*}) \right]^* $$
dual to the above $\delta^\prime$ right coaction\footnote{ Remember 
that in the ``classical'' case (\ie real forms of Lie algebras and 
their enveloping algebras), $(Sh)^{*}=h$ for the Lie algebra 
generators, and we recognize the usual equation $\bar\rho = {*} \rho 
{*}$ defining the conjugate representation. }.
Therefore the compatibility relation (\ref{Hopf_star_on_module}) 
between the stars on $H$ and $V$
can also be viewed as a very particular case of equivalency of 
representations: $\rho$ and $\bar\rho$ should just coincide. Given 
the star operations, a representation $\rho$ is called {\it 
selfconjugate} if there exists an invertible operator $U:V\mapsto V$ 
such that
$$
U^{-1} \rho[h] U = \bar\rho[h] \ .
$$

Up to now we did not assume that the representation space $V$ was 
endowed with a scalar product $(\cdot\,,\cdot)$. Therefore, we can 
not impose, at this point, that $U$ should be unitary. We can not 
assume, either, that the star operation on $V$ is an antiunitary 
operator, $(v^{*},w^{*})=(w,v)$. For the same reason too, the 
notation $\dag$ (adjoint) was avoided. In any case, a Hopf algebra 
is, in particular, an associative algebra, and if it is so happens 
that a {\sl real} Hopf algebra $H_{\RR}$ can be defined the usual 
classification for representations of real associative algebras on 
complex Hilbert spaces will, of course, also hold. We could have 
three types of representations, complex, real, and quaternionic; we 
refer the reader to standard textbooks (see for instance 
\cite{Bourbaki}\cite{Simon}).

%%%%%%%%%%%%%%%%%%%%%%%%%%%%%%%%%%%%%%%%%%%%%%%%%%%%%%%%%%%%%%%%%%%%%%%%%% 
\subsubsection{Compatible stars on module-algebras} 

Instead of a comodule $V$, we now take a right $F$-comodule algebra 
$M$, \ie we assume that the right coaction $\delta_{R}$ is an algebra 
homomorphism from $M$ to $M\otimes F$,
$$
\delta_{R}(zw)=\delta _{R} z \, \delta _{R} w \ .
$$
The map $\delta^\prime: M \rightarrow M \otimes F$ defined as above 
will again be an algebra homomorphism, \ie 
$\delta^\prime(zw)=\delta^\prime z \, \delta^\prime w$. 
Thus equation (\ref{Hopf_star_on_comodule}) is still a good 
requirement when the comodule $M$ is an algebra and shows that 
compatibility of the coaction with a given Hopf star operation needs 
only to be verified on the (algebra) generators\footnote{
In the case of a module algebra, the star operation is of course 
assumed to be antimultiplicative ($(xy)^{*} = y^{*} x^{*}$).
}.

Obviously the dual equation (\ref{Hopf_star_on_module}) defining 
compatibility of Hopf stars on left modules will also have the same 
properties. Remember that, being a right $F$-comodule algebra, $M$ 
supports a left action of the dual $H$ of $F$ and indeed this action 
is compatible with the product in $M$ (call 
$\Delta h = h_{1}\otimes h_{2}$):
$$
h \triangleright (zw) = 
   (h_1\triangleright z)(h_2\triangleright w) \ .
$$

%%%%%%%%%%%%%%%%%%%%%%%%%%%%%%%%%%%%%%%%%%%%%%%%%%%%%%%%%%%%%%%%%%%%%%%%%% 
\subsubsection{Example of the reduced $SL_{q}(2,\CC)$ at $q^N = 1$} 
\label{s:stars_on_FHM}

\medskip

\noindent
{\bf Hopf stars on ${\mathcal F}$ and ${\mathcal H}$} 

\medskip

\noindent
First of all, remember that in the quantum case one has three 
possibilities for the star operations on $Fun(SL_q(2,\CC))$ (up to 
star-Hopf homomorphisms). Given the conventions chosen in 
\cite{CoGaTr-e}, they are given on generators by: 

\begin{itemize}
\item
The real form $Fun(SU_q(2))$: \quad $a^* = d$, $b^* = -qc$, 
$c^* = -q^{-1} b$ and $d^* = a$. Moreover, $q$ should be real. 

\item
The real form $Fun(SU_q(1,1))$: \quad $a^* = d$, $b^* = qc$, 
$c^* = q^{-1} b$ and $d^* = a$. Moreover, $q$ should be real. 

\item
The real form $Fun(SL_q(2,\RR))$: the conjugation is given by
\begin{eqnarray}
a^* &=& a \nonumber \\
b^* &=& b \label{Hopf_star_on_F} \\
c^* &=& c \nonumber \\
d^* &=& d \ . \nonumber
\end{eqnarray}
Here $q$ can be complex but it should be a phase.

\end{itemize}

\noindent
When $q=\pm i \;$ ---hence $q^{4}=1$--- there are still two other 
Hopf star structures that have no classical limit (see 
\cite{Chari-Pressley} and references therein). A systematic analysis 
of real forms for special linear quantum groups $SL_{q}(n)$ was made 
by \cite{Jain-Ogievetsky} and, in the case of $GL_{p,q}(2)$ or 
$GL_{\alpha}^J(2)$, by \cite{Ewen-Ogievetsky}. 

It is already clear from these results that taking $q$ a root of 
unity is incompatible with the $SU_q$ and $SU_{q}(1,1)$ real forms. 
The only possibility if we assume $q^N = 1$ is to choose the Hopf 
star corresponding to $Fun(SL_q(2,\RR))$. Moreover, in such a case 
the star is compatible with the finite dimensional Hopf algebra 
quotient $\mathcal{F}$ obtained by factoring this quantum group by 
the Hopf ideal defined by \cite{CoGaTr-e} 
$a^{N}=d^{N} = \one \,,\ b^{N}=c^{N}=0$
(take $N$ odd here, and $q$ a primitive $N$-th root of unity). 

The corresponding dual star on the dual Hopf algebra $U_q(sl(2,\CC))$ 
(see \cite{CoGaTr-e} or \ref{a:structure_of_H} for its structure) is 
\begin{eqnarray}
X_{+}^{*} &=& -q^{-1}\,X_{+} \nonumber \\
X_{-}^{*} &=& -q\,X_{-} \label{Hopf_star_on_H} \\
K^{*} &=& K \ . \nonumber
\end{eqnarray}

\noindent
Here one can also factor the quantum enveloping algebra by the Hopf 
ideal defined by $K^{N}=\one\,,\ X_{+}^{N}=0\,,\ X_{-}^{N}=0$ and the 
same remarks concerning the fact that the stars passes to the 
quotient $\mathcal H$ apply \cite{CoGaTr-e}.

\medskip

\noindent
{\bf Compatible star on the quantum plane ${\mathcal M}$} 

\medskip

\noindent
The quantum group $Fun(SL_q(2,\CC))$ coacts on the quantum plane 
algebra generated by $x,y$ such that $xy=qyx$. For a root of unity 
this algebra can be quotiented by the ideal defined by 
$x^N=y^N=\one$, to obtain a finite dimensional algebra that we call 
$\mathcal M$. $\mathcal{M}$ is a right comodule algebra for 
$\mathcal{F}$ and the right coaction is given by
$\delta_{R} \left( \begin{array}{cc} x & y \end{array} \right) = 
\left( \begin{array}{cc} x & y \end{array} \right) \dot{\otimes} 
\left( \begin{array}{cc} a & b \\ c & d \end{array} \right)$. 
The reduced quantized universal enveloping algebra $\mathcal{H}$ acts 
on this quantum plane (for compatible formulae for actions and 
coactions, see for instance \cite{CoGaTr-e}). Up to equivalences (now 
$*$-homomorphisms) there is only one conjugation on this quantum 
plane compatible with the requirements (\ref{Hopf_star_on_comodule}) 
or (\ref{Hopf_star_on_module}). It works for both the infinite 
dimensional algebra or its reduced (finite) quotients when $q^N = 1$. 
It is
\begin{eqnarray}
x^{*}&=&x \cr
y^{*}&=&y \ . \nonumber
\end{eqnarray}
Notice that although the star is the identity on the generators it is 
non-trivial on $\mathcal{M}$ since it is an antimultiplicative 
operation and, for instance, $(xy)^{*} = q^{-1} xy$. 

%%%%%%%%%%%%%%%%%%%%%%%%%%%%%%%%%%%%%%%%%%%%%%%%%%%%%%%%%%%%%%%%%%%%%%%%%% 
%%%%%%%%%%%%%%%%%%%%%%%%%%%%%%%%%%%%%%%%%%%%%%%%%%%%%%%%%%%%%%%%%%%%%%%%%% 
\subsection{Twisted Hopf stars}
\label{s:twisted_star}

As we mentioned before there is an alternative way of relating the 
Hopf and star structures on a Hopf algebra. It reduces to replacing 
in (\ref{normal_Hopf_star}) the equation for the coproduct 
by\footnote{ 
It could even be written $\Delta * = *_{_{op}} \Delta$ at the expense 
of using a flipped definition of the star on the tensor product:
$*_{_{op}} (f\otimes g) = g^* \otimes f^*$.
}

\begin{eqnarray*}
   \Delta * &=& (* \otimes *) \Delta^{op} \ , 
\end{eqnarray*} 

\noindent
Given such a twisted star on a Hopf algebra $H$, the dual Hopf 
algebra $F=H^\star$ can be also endowed with a dual twisted Hopf 
star. One just has to define it by

\begin{equation}
    \langle f^* \,,h\rangle =\overline{\langle f\,,h^*\rangle} \ . 
\label{twisted_dual_*}
\end{equation}

\noindent
It can be readily verified that this operation is a twisted Hopf star 
on $F$. As in the untwisted case, a relation involving the antipode 
is automatically fulfilled. Now the antipode and the star commute, 

\begin{eqnarray}
   S \, * &=& * \, S \ .
\end{eqnarray}

\noindent
This is so because $*S*$ is again an antipode, which is unique. 

%%%%%%%%%%%%%%%%%%%%%%%%%%%%%%%%%%%%%%%%%%%%%%%%%%%%%%%%%%%%%%%%%%%%%%%%%% 
\subsubsection{Compatible twisted stars on modules} 

Let $V$ be again a right $F$-comodule. Given $*_F$ a twisted Hopf 
star on $F$ we would now like to use it to restrict the possible 
choices for a star $*_V$ on $V$, as it was done with equation 
(\ref{Hopf_star_on_comodule}) in the pure Hopf case. 

$*_F$ being twisted, 
$(* \otimes *)\, \delta_R \, * : V \mapsto V\otimes F$ is no longer 
a right coaction, however 
$\tau (* \otimes *) \,\delta_R \,* : V \mapsto F\otimes V$ is a 
left one. Moreover
$(\id \otimes S) (* \otimes *) \, \delta_R \,* = 
      (* \otimes *) (\id \otimes S) \, \delta_R \,*$ 
is again a right coaction. Consequently we may require

\begin{equation}
(\id \otimes S) \delta_R (z^*) = \left( \delta_R z \right)^* \ , 
\qquad z \in V \ ,
\label{twisted_star_on_comodule_space}
\end{equation}
or the following dual expression for the corresponding action of $H$ 
on the module $V$:

\begin{equation}
   h \triangleright z^* = \left[ (Sh)^* \triangleright z \right]^* \ , 
      \qquad z \in V \ .
\label{Hopf_star_on_module_space}
\end{equation}
Notice that this condition looks formally like 
(\ref{Hopf_star_on_module}).

%%%%%%%%%%%%%%%%%%%%%%%%%%%%%%%%%%%%%%%%%%%%%%%%%%%%%%%%%%%%%%%%%%%%%%%%%% 
\subsubsection{Compatible twisted stars on module-algebras} 

If we now let $V$ be an $F$-comodule {\em algebra} (we then call it 
$M$ rather than $V$), it happens that
(\ref{twisted_star_on_comodule_space}) is not a reasonable condition 
anymore, because $(\id \otimes S) \delta_R *$ and $* \delta_R$ have 
different homomorphism behaviour. It may also be said that $(\id 
\otimes S) (* \otimes *) \delta_R *$ is not an R-coaction {\em on an 
algebra} but only a coaction; it doesn't preserve the product on $M$. 

As $\tau \, (* \otimes *)\, \delta_R \, *$ is a good homomorphism, 
the way out to constrain $*_M$ is to choose some {\em other} left 
algebra-coaction $\delta_L$ on $M$ and impose 

\begin{equation}
\delta_R (z^*) = \left( \delta_L z \right)^{*_{op}} \ , 
        \qquad z \in M \ ,
\label{twisted_star_on_comodule_algebra} 
\end{equation} 

\noindent
where now the star $*_{_{op}}$ on the right hand side includes the 
tensorial flip (on $F \otimes M$ it is given by 
$*_{_{op}}(f \otimes z) = z^* \otimes f^* \,, \ z \in M, \ f \in F$). 
Remark that for many interesting cases we have both natural left and 
right coactions; this is for instance the case for quantum planes. 

The dual condition involves the left and right actions of $H$ on $M$ 
which are dual to $\delta_R$ and $\delta_L$, they are respectively 
denoted by $\triangleright$ and $\triangleleft$. It reads:

\begin{equation}
z^* \triangleleft h = \left[ h^* \triangleright z \right]^* \ , 
\qquad h\in H\, , \ z\in M \ .
\label{twisted_star_on_module_algebra}
\end{equation}

%%%%%%%%%%%%%%%%%%%%%%%%%%%%%%%%%%%%%%%%%%%%%%%%%%%%%%%%%%%%%%%%%%%%%%%%%% 
\subsubsection{Example of the reduced $SL_{q}(2,\CC)$ at $q^N = 1$} 

\medskip

\noindent
{\bf Twisted Hopf stars on ${\mathcal F}$ and ${\mathcal H}$} 

\medskip

\noindent
On both the reduced and unreduced $SL_{q}(2,\CC)$, the twisted stars 
are essentially the following\footnote{
The operation defined on generators by
$a^{*}=d, d^{*}=a, b^{*}=\pm b$ and $c^{*}=\pm c$ ``almost works'', 
in the sense that it defines a twisted star in $GL_{q}(2,\CC)$ but it 
is incompatible with the determinant condition defining 
$SL_{q}(2,\CC)$.}
(\ie up to automorphisms):

\begin{eqnarray}
a^*&=&a \nonumber \\ b^*&=&\pm c \label{twisted_star_on_F} \\
c^*&=&\pm b \nonumber \\ d^*&=&d \nonumber \ . 
\end{eqnarray}

\noindent So we have two of them, and the corresponding dual twisted 
stars are given by

\begin{eqnarray}
X_{+}^* &=& \pm X_{-} \nonumber \\
X_{-}^* &=& \pm X_{+} \label{twisted_star_on_H} \\
K^* &=& K^{-1} \nonumber \ .
\end{eqnarray}

\noindent
Thus we see that, when $q$ is a root of unity, these twisted stars 
allow one to recover
the $SU(2)$ ($+$ sign) and $SU(1,1)$ ($-$ sign) real forms, something 
that would be otherwise forbidden with a {\it true} Hopf star 
operation.

\medskip

\noindent
{\bf Compatible star on the quantum plane ${\mathcal M}$} 

\medskip

\noindent
On the quantum plane there is, again up to equivalence, only one star 
structure
compatible in the sense (\ref{twisted_star_on_comodule_algebra}) or 
(\ref{twisted_star_on_module_algebra}) with each of the twisted stars 
(\ref{twisted_star_on_F}) or (\ref{twisted_star_on_H}). These twisted 
stars are respectively given by

\begin{eqnarray}
x^{*} &=& x \label{twisted_star_on_M} \\
y^{*} &=& \pm y \ . \nonumber
\end{eqnarray}

%%%%%%%%%%%%%%%%%%%%%%%%%%%%%%%%%%%%%%%%%%%%%%%%%%%%%%%%%%%%%%%%%%%%%%%%%% 
%%%%%%%%%%%%%%%%%%%%%%%%%%%%%%%%%%%%%%%%%%%%%%%%%%%%%%%%%%%%%%%%%%%%%%%%%% 
\subsection{Stars and tensor products}

\subsubsection{Tensor product of matrices} 

If
$$
m = \pmatrix a & b \cr c & d \endpmatrix \qquad \text{and} \qquad 
M = \pmatrix A & B \cr C & D \endpmatrix 
$$
are two matrices with {\sl non commutative} entries belonging to a 
ring $\mathcal B$, then it is standard to define their tensor product 
as 
$$
m \otimes M \doteq \pmatrix
aA & aB & bA & bB \cr
aC & aD & bC & bD \cr
cA & cB & dA & dB \cr
cC & cD & dC & dD \endpmatrix \ .
$$
We now define a different tensor product, $\otimes_{op}$, by 
$$
M \otimes_{op} m = \pmatrix
Aa & Ba & Ab & Bb \cr
Ca & Da & Cb & Db \cr
Ac & Bc & Ad & Bd \cr
Cc & Dc & Cd & Dd \endpmatrix \ ,
$$
the difference being that now the matrix which determines the coarse 
structure of the tensor product is the second one. It is clear and 
well known that $m \otimes M \neq M \otimes m$, independently of 
whether $\mathcal B$ is commutative or not. However, we see that when 
$\mathcal B$ is abelian, $m \otimes M = M \otimes_{op} m$. The 
previous calculation tells us how to modify this result when 
$\mathcal B$ is not commutative: calling $\mathcal B^{op}$ the same 
ring with {\em opposite} multiplication (so that 
$A\, ._{op} \, a = a.A$, for example), we obtain
$$
m(\mathcal B) \otimes M(\mathcal B) = 
M(\mathcal B^{op}) \otimes_{op} m(\mathcal B^{op})
$$
where the notation 
$M(\mathcal B^{op}) \otimes_{op} m(\mathcal B^{op})$ means that we 
first take the opposite tensor product of the two matrices and 
subsequently we multiply the matrix elements in the opposite order. 

Suppose in addition that the ring $\mathcal B$ is endowed with a star 
operation~$*$, and call $\dag$ the conjugation of matrices with 
$\mathcal B$-entries. In the case of $2\times 2$ matrices, this reads
$$
{\pmatrix a & b \cr c & d \endpmatrix}^\dag \doteq 
     \pmatrix a^{*} & c^{*} \cr b^{*} & d^{*} \endpmatrix \ .
$$
So defined $\dag$ is antimultiplicative. Moreover, direct calculation 
shows that
$$
(m \otimes M)^\dag = M^\dag \otimes_{op} m^\dag \ .
$$
When $\mathcal B$ is commutative, the previous right hand side can be 
written simply $m^\dag \otimes M^\dag$.

%%%%%%%%%%%%%%%%%%%%%%%%%%%%%%%%%%%%%%%%%%%%%%%%%%%%%%%%%%%%%%%%%%%%%%%%%% 
\subsubsection{Tensor product of representations} 

Now let $\mathcal A$ be an algebra. Take $\rho_{1}$ and $\rho_{2}$ 
two representations of $\mathcal A$ in vector spaces $V_1$ and $V_2$. 
Then, once bases are chosen, $\rho_{1}(a)$ and $\rho_{2}(a)$, with 
$a\in {\mathcal A}$, are two matrices with commutative entries. 

It is clear that $\rho_{1} \otimes \rho_{2}$ is a representation of 
the algebra ${\mathcal A}\otimes {\mathcal A}$, indeed, with 
$a\otimes b \in {\mathcal A}\otimes {\mathcal A}$, we have 
$$
[\rho_{1} \otimes \rho_{2}](a\otimes b) = 
           \rho_{1}(a) \otimes \rho_{2}(b) \ .
$$
However, this is {\em not} a representation of ${\mathcal A}$, unless 
we have a coproduct (algebra homomorphism) from ${\mathcal A}$ to 
${\mathcal A}\otimes {\mathcal A}$: using 
$$
a \in {\mathcal A} \rightarrow \Delta a \doteq a_{1}\otimes a_{2} 
\in {\mathcal A} \otimes {\mathcal A}
$$
one defines $\rho_{1} \otimes \rho_{2}$ as a representation of 
${\mathcal A}$ by setting
$$
[\rho_{1} \otimes \rho_{2}][a] \doteq
[\rho_{1} \otimes \rho_{2}](\Delta a) \ .
$$
If $\mathcal A$ is a Hopf algebra, we are in such a situation. This 
is what we assume from now on.

Now suppose that $\mathcal A$ has a star operation, and that 
$(\rho_{1},\dag)$, $(\rho_{2},\dag)$ are star representations of this 
Hopf algebra on modules $V_1$, $V_2$ (each one endowed with a scalar 
product for which the adjoint is denoted by $\dag$). So, we have 
$$
\rho_{1}(u^{*}) = (\rho_{1} (u))^\dag \qquad \mbox{and} \qquad 
\rho_{2}(u^{*}) = (\rho_{2} (u))^\dag \ .
$$
We shall now suppose that the star is, somehow, compatible with the 
Hopf structure. We shall discuss the Hopf star and twisted Hopf star 
cases. 

We first suppose that $*$ is a Hopf star. It then commutes with 
$\Delta$, and
\begin{eqnarray*}
[\rho_{1} \otimes \rho_{2}][a^{*}] &=&
[\rho_{1} \otimes \rho_{2}](\Delta a^{*}) = 
[\rho_{1} \otimes \rho_{2}](* \Delta a) = 
[\rho_{1} \otimes \rho_{2}](a_{1}^{*}\otimes a_{2}^{*}) \cr 
{} &=& \rho_{1}(a_{1}^{*}) \otimes \rho_{2}(a_{2}^{*}) = 
(\rho_{1}(a_{1}))^\dag \otimes (\rho_{2}(a_{2}))^\dag = 
(\rho_{1}(a_{1}) \otimes \rho_{2}(a_{2}))^\dag \cr 
{} &=& ([\rho_{1} \otimes \rho_{2}](a_{1}\otimes a_{2}))^\dag = 
([\rho_{1} \otimes \rho_{2}](\Delta a))^\dag \cr
{} &=& ([\rho_{1} \otimes \rho_{2}][a])^\dag
\end{eqnarray*}
Therefore, $\rho_{1}\otimes \rho_{2}$ is also a $*$-representation. 

We now suppose that $*$ is a twisted Hopf star. It no longer commutes 
with $\Delta$ but intertwines it with the opposite coproduct 
$\Delta^{op}$. In this case
\begin{eqnarray*}
[\rho_{1} \otimes \rho_{2}][a^{*}] &=&
[\rho_{1} \otimes \rho_{2}](\Delta a^{*}) = 
[\rho_{1} \otimes \rho_{2}](* \Delta^{op} a) = 
[\rho_{1} \otimes \rho_{2}](a_{2}^{*}\otimes a_{1}^{*}) \cr
 {} &=& \rho_{1}(a_{2}^{*}) \otimes \rho_{2}(a_{1}^{*}) = 
(\rho_{1}(a_{2}))^\dag \otimes (\rho_{2}(a_{1}))^\dag = 
(\rho_{1}(a_{2}) \otimes \rho_{2}(a_{1}))^\dag \cr 
{} &=& ([\rho_{1} \otimes \rho_{2}](a_{2}\otimes a_{1}))^\dag = 
([\rho_{1} \otimes \rho_{2}](\Delta^{op} a))^\dag \cr
{} &\neq& ([\rho_{1} \otimes \rho_{2}][a])^\dag 
\end{eqnarray*}

\noindent
Therefore, $\rho_{1}\otimes \rho_{2}$ is {\sl not} a $*$-representation 
for a twisted $*$. However, we have the possibility of defining ``another'' 
tensor product of representations\footnote{ 
When $\mathcal A$ is quasitriangular, we recall that the two 
coproducts are related by an $R$-matrix as follows: 
$\Delta^{op}(a) = R \Delta(a) R^{-1}$.
},
called $\otimes_{op}$, as follows:
$$
[\rho_{1}\otimes_{op} \rho_{2}][a] \doteq 
[\rho_{1}\otimes \rho_{2}](\Delta^{op} a) \ .
$$
With this notation at hand, we can write 
$$
[\rho_{1}\otimes \rho_{2}][a^{*}] =
([\rho_{1}\otimes_{op} \rho_{2}][a])^\dag \ .
$$

For this reason, ``true'' Hopf stars are usually preferred in mathematics, 
as the category of $*$-representations is closed under tensor product.
Another possibility, the one employed in CFT's, is to truncate tensor
products (see Section~\ref{s:discussion}). Star representations are 
closed under this truncated tensor product ---for both types of stars.

%%%%%%%%%%%%%%%%%%%%%%%%%%%%%%%%%%%%%%%%%%%%%%%%%%%%%%%%%%%%%%%%%%%%%%%%%% 
\subsubsection{Hopf action on vectors with non commutative elements} 

We now suppose that $\rho_{1}$ and $\rho_{2}$ are no longer complex 
matrices but matrices with elements taken in a star algebra $\mathcal 
B$. We still assume that we have a left action, in the sense 
$\rho_{i}(ab) = \rho_{i}(a) \rho_{i}(b)$, but this is not a 
representation in the usual sense. As before we assume that $\mathcal 
A$ is endowed with a star operation and that $(\rho_{i},\dag)$ are 
star representations, in the sense $\rho(a^*) = (\rho(a))^\dag$, 
where $\dag$ transposes the matrix $\rho(a)$ and takes the conjugate 
(in $\mathcal B$) of each element.

If we suppose that $*_\mathcal{A}$ is a Hopf star, then a direct 
calculation shows that
$$
[\rho_{1}\otimes \rho_{2}][a^{*}] =
([\rho_{1}^{op} \otimes \rho_{2}^{op} ][a])^\dag \ .
$$
Usually, for $\mathcal{B} = \CC$, we have $\rho^{op} = \rho$, but 
this is not so in general; the upper index ``op'' in $\rho^{op}(a)$ 
reminds us that we should use the opposite multiplication of 
$\mathcal B$ when making product of matrices such as $\rho^{op}(a)$. 

If we take instead a twisted Hopf star, the conclusion is now:
$$
[\rho_{1}\otimes \rho_{2}][a^{*}] =
([\rho_{1}^{op} \otimes_{op} \rho_{2}^{op} ][a])^\dag \ .
$$

%%%%%%%%%%%%%%%%%%%%%%%%%%%%%%%%%%%%%%%%%%%%%%%%%%%%%%%%%%%%%%%%%%%%%%%%%% 
%%%%%%%%%%%%%%%%%%%%%%%%%%%%%%%%%%%%%%%%%%%%%%%%%%%%%%%%%%%%%%%%%%%%%%%%%% 
%%%%%%%%%%%%%%%%%%%%%%%%%%%%%%%%%%%%%%%%%%%%%%%%%%%%%%%%%%%%%%%%%%%%%%%%%% 

\section{Invariant scalar products}
\label{s:invariant_scalar_products}

\subsection{Compatibility with Hopf stars} 
\label{s:invariant_scalar_product_for_Hopf}

Defining the notion of an invariant scalar product $(\cdot,\cdot)$ on 
a representation space $V$ of a quantum group $H$ is not as 
straightforward as in the classical case. We want the scalar product 
to commute with the action
of the Hopf algebra, in the appropriate sense. However, in order to 
get a relation which needs to be checked only on the quantum group 
generators, we want this condition to be a (linear) homomorphism in 
the $H$ variable. Given that the scalar product is antilinear in one 
of its variables, there are two ways of achieving this\footnote{ In 
the ``classical case'' (real form of some Lie algebra), both formulae 
read $(z,h \triangleright w) + (h \triangleright z, w) = 0$.}, 

\begin{equation}
\epsilon(h) (z,w) =
((*\,S\, h_1) \triangleright z , h_2 \triangleright w) 
\label{invariant_scalar_product}
\end{equation}
or
\begin{equation}
\epsilon(h) (z,w) =
((S* h_1) \triangleright z , h_2 \triangleright w) \ . 
\label{invariant_scalar_product_2}
\end{equation}

\noindent
We refer the reader to \cite{CoGaTr-l} for a more detailed discussion.

As the Hopf star doesn't commute with the antipode, since 
$S \, * = * \, S^{-1}$, (\ref{invariant_scalar_product}) and 
(\ref{invariant_scalar_product_2}) are, in general, two different 
conditions.

For the scalar product to be invariant in the sense of equation 
(\ref{invariant_scalar_product}), one only needs the quantum group 
action to be given by a $*$-{\sl representation}\footnote{ If the
action of $h$ is implemented by a linear operator $\rho[h]$ on $V$, 
this condition simply reads $\rho[h^{*}]= (\rho[h])^\dag$ where 
$\dag$ is the usual adjoint operator.}:

\begin{equation}
(h \triangleright z , w) = (z , h^* \triangleright w) \ . 
\label{star_representation}
\end{equation}

\noindent
Notice that (\ref{star_representation}) implies 
(\ref{invariant_scalar_product}) but not conversely. In the same way 
the alternative requirement
$(h \triangleright z , w) = (z , S^2(h^*) \triangleright w)$ implies 
that condition (\ref{invariant_scalar_product_2}) is satisfied. 
However, in our examples, we will choose to work with 
$*$-representations, and therefore with invariant scalar products in 
the sense (\ref{invariant_scalar_product}). 

Assuming a non-degenerate pairing between $H$ and its dual $F$, and 
extending the notation $(\cdot,\cdot)$ to the following $F$-valued 
sesquilinear map on $V\otimes F$
$$
(v \otimes f,w \otimes g) \doteq (v,w) f^* g \ , \qquad v,w\in V, \ 
f,g \in F
$$
we may write the previous equations in the dual picture in a very 
simple way. The first invariance condition reads
\begin{equation}
(\delta_R \, v, \delta_R \, w) = (v,w) \one_F \ , \nonumber 
\end{equation}
whereas the $*$-representation requirement 
(\ref{star_representation}) reads
\begin{equation}
(v,\delta_R w) = ((\id \otimes S)\delta_R v,w) \ . \nonumber 
\end{equation}
Again, this latter requirement implies the former. 

Now let $\{ v_i \}$ be a basis of the vector space $V$, and call 
$G_{ij} = (v_i,v_j)$ the corresponding metric. Moreover, define the 
matrix of $h \in H$ in such a basis by  
$h \triangleright v_i \doteq ||h||_{ji} v_j$.
{}From equation (\ref{star_representation}) it is now trivial to get 
the matrix identities
\begin{equation}
   ||h||^\dag G = G ||h^*|| \ ,
\label{star_representation_in_basis}
\end{equation}
where $\dag$ denotes the transposed conjugate matrix. In particular, for 
an orthonormal basis this reduces to $||h||^\dag = ||h^*||$.

%%%%%%%%%%%%%%%%%%%%%%%%%%%%%%%%%%%%%%%%%%%%%%%%%%%%%%%%%%%%%%%%%%%%%%%%%% 
%%%%%%%%%%%%%%%%%%%%%%%%%%%%%%%%%%%%%%%%%%%%%%%%%%%%%%%%%%%%%%%%%%%%%%%%%% 

\subsection{Compatibility with twisted Hopf stars} 

The previous discussion 
(Section~\ref{s:invariant_scalar_product_for_Hopf}) does not use the 
fact that the chosen star should be a ``true'' Hopf star operation; 
therefore, the same invariance conditions 
(\ref{invariant_scalar_product}) and (\ref{invariant_scalar_product_2}) 
still apply in the twisted case. However now $S*=*S$, so that both 
conditions coincide.

The invariance requirement is still automatically satisfied if the 
representation of $H$ under study is a $*$-representation (formula 
(\ref{star_representation})). However, now the dual formulas are 
slightly different, due to the absence of the antipode in the duality 
(\ref{twisted_dual_*}). The scalar product will be called invariant if 

\begin{equation}
((\id\otimes S)\delta_R \, v, \delta_R \, w) = (v,w) \one_F \ , \nonumber 
\end{equation}
and the (co)representation will be a $*$-(co)representation if 
\begin{equation}
(v,\delta_R w) = (\delta_R v,w) \ . \nonumber 
\end{equation}

\noindent
Selecting a basis of $V$ we can write, exactly as in the untwisted case,
$||h||^\dag G = G ||h^*||$ for any $h \in H$.

%%%%%%%%%%%%%%%%%%%%%%%%%%%%%%%%%%%%%%%%%%%%%%%%%%%%%%%%%%%%%%%%%%%%%%%%%% 
%%%%%%%%%%%%%%%%%%%%%%%%%%%%%%%%%%%%%%%%%%%%%%%%%%%%%%%%%%%%%%%%%%%%%%%%%% 

\subsection{Quantum metric and quantum symplectic form on 
            $M(2,\mathcal{F})$} 

\medskip

\noindent
{\bf Untwisted case}

\medskip

The $q$-deformed symplectic form in two dimensions (one may call it 
the $q$-deformed epsilon tensor) is given by the matrix $$
\Sigma \doteq \left( \begin{array}{cc} 0 & q^{-1/2} \\ 
-q^{1/2} & 0 \end{array} \right) \ .
$$

\noindent
In fact the $*$-representation condition implies for the true Hopf 
star case the equation
$$
T^\dag \, \Sigma \, T = \Sigma \ .
$$
Here
$$
T \doteq \left( \begin{array}{cc} a & b \\ c & d \end{array} \right) 
$$
is the multiplicative matrix of generators of the quantum function 
group $SL_{q}(2,\CC)$, and the $\dag$ operation corresponds to 
applying $*$ to the elements and transposing the matrix: 
$$
T^\dag \doteq \left( \begin{array}{cc} a^* & c^* \\ b^* & d^* 
                     \end{array} \right) \ .
$$
Notice that the above equation is different from 
(\ref{star_representation_in_basis}) as there is a duality involved, 
there $h \in H, \: ||h||_{ij} \in \CC$, whereas here $T_{ij} \in F$.
Using the star (\ref{Hopf_star_on_F}) corresponding to $SL_{q}(2,\RR)$ 
and fixing a global factor by requiring hermiticity of $\Sigma$, we 
finally obtain the ``invariant metric'' given above. 

\medskip

\noindent
{\bf Twisted case}

\medskip

\noindent
Now, as the duality between the star on a Hopf algebra and its dual 
differs from the one in the untwisted case, the $*$-representation 
condition implies the relation
$$
(ST)^\dag \, \Sigma \, T = \Sigma \ ,
$$
where $S$ is the antipode. Taking the twisted conjugacy
$a^*=a\,,\ b^*=\pm c\,,\ c^*=\pm b$ and $d^*=d$ we get the metric
$$
\Sigma_\pm \doteq
\left( \begin{array}{cc} 1 & 0 \\ 0 & \pm 1 \end{array} \right) \ , 
$$
as we would expect in a (twisted) $SU(2)$ and $SU(1,1)$ case, 
respectively.

%%%%%%%%%%%%%%%%%%%%%%%%%%%%%%%%%%%%%%%%%%%%%%%%%%%%%%%%%%%%%%%%%%%%%%%%%% 
%%%%%%%%%%%%%%%%%%%%%%%%%%%%%%%%%%%%%%%%%%%%%%%%%%%%%%%%%%%%%%%%%%%%%%%%%% 

\subsection{Invariant scalar products for $\mathcal H$ endowed with a 
            Hopf star} 

\subsubsection{Invariant scalar products on the indecomposable 
               representations of $\mathcal H$}

This was worked out in Appendix E of \cite{CoGaTr-e}. Here we repeat
the expressions for the matrices of scalar products $G$ just for 
completeness and to ease the comparison with the twisted case.
Technically this is done by solving the set of linear equations 
(\ref{star_representation_in_basis}) for the coefficients $G_{ij}$
taking $h=X_{\pm}$ and $K$, and imposing hermiticity of $G$. Each entry 
in the list below corresponds to an indecomposable representation,
remember that $3_{irr}$ is projective and irreducible, whereas $6_{odd}$ 
and $6_{eve}$ are projective indecomposable (with corresponding 
irreducible representations of dimensions $1$ and $2$, respectively). We 
only single out the following salient features (notice that $G$ is 
always given up to an overall normalization factor):

\medskip

\begin{itemize}

\item[$\bullet \: \mathbf{3_{irr}}$]
   We get
   {\small
   $$
      G = \pmatrix 0& 0& -q^2\cr 0& 1& 0\cr -q& 0& 0
          \endpmatrix \qquad \text{and}
      \qquad \sigma = (++-) \ .
   $$}
   
   \noindent
   The index of $G$ (maximal dimension of each of the two maximally 
   isotropic subspaces) is therefore $1$, and the Witt decomposition 
   reads $3 = 1 + 1 + 1$. 
   \medskip

\item[$\bullet \: \mathbf{6_{odd}}$]
   With $\beta \in \RR$ we have here
   {\small
   $$
      G = \pmatrix    0&   0& 0& q&     0& 0\cr
                      0&   0&-q& 0&     0& 0\cr
                      0&-q^2& 0& 0&     0& 0\cr
                    q^2&   0& 0& 0&     0& 0\cr
                      0&   0& 0& 0& \beta& 1\cr
                      0&   0& 0& 0&     1& 0 \endpmatrix \: 
                      \sim \diag(1,1,-1,-1,\lambda_+,\lambda_-) \ ,
   $$}
   
   \noindent
   with $\lambda_+ > 0, \lambda_- < 0$. $G$ is neutral, as its signature 
   is $\sigma = (+++ - - -)$. The index of $G$ is $3$ and the Witt 
   decomposition reads $6 = 3 + 3$.
   \medskip

\item[$\bullet \: \mathbf{5_{odd}}$]
   Taking $\beta,\gamma \in \RR$, $g \in \CC$,
   {\small
   $$
      G = \pmatrix  0& 0&    iq\gamma&       g& 0\cr
                    0& 0& -q^2 \bar g& iq\beta& 0\cr
                    -i q^2 \gamma&         -qg& 0& 0& 0\cr
                           \bar g& -iq^2 \beta& 0& 0& 0\cr
                                0&           0& 0& 0& 0 \endpmatrix \:
          \sim \diag(\lambda_+,-\lambda_+,\lambda_-,-\lambda_-,0)
   $$}
   
   \noindent
   and $\sigma = (++--0)$.
   \medskip

\item[$\bullet \: \mathbf{3_{odd}}$]
   Now
   {\small
   $$
      G = \pmatrix 0& iq& 0\cr -iq^2& 0& 0\cr 0& 0& 0 \endpmatrix 
      \qquad \text{and} \qquad \sigma = (+-0) \ .
   $$}
   \medskip

\item[$\bullet \: \mathbf{6_{eve}}$]
   The metric should be ($\beta \in \RR$)
   {\small
   $$
      G = \pmatrix  0& 0& iq\beta& -iq& 0& 0\cr
                    0& 0&     -iq&   0& 0& 0\cr
                    -iq^2 \beta& iq^2& 0& 0& 0& 0\cr
                           iq^2&    0& 0& 0& 0& 0\cr
                    0& 0& 0& 0&  0& i\cr
                    0& 0& 0& 0& -i& 0 \endpmatrix \ ,
   $$}
   
   \noindent
   with a signature $\sigma = (+++---)$ for any $\beta$. As in 
   the $6_{odd}$ case, $G$ is neutral with an index of $3$, and 
   the Witt decomposition reads $6 = 3 + 3$.
   \medskip

\item[$\bullet \: \mathbf{4_{eve}}$]
   Having $\alpha, \beta \in \RR$, and $g \in \CC$ we may write
   {\small
   $$
      G = \pmatrix 0& 0&      0&     0\cr
                   0& 0&      0&     0\cr
                   0& 0& \alpha&     g\cr
                   0& 0& \bar g& \beta \endpmatrix 
   $$}
   
   \noindent
   Now its signature obviously depends on the parameters.
   \medskip

\item[$\bullet \: \mathbf{3_{eve}}$]
   Here we have simply $G = \diag( 0, 0, 1)$, and $\sigma = (+00)$.
   \medskip

\item[$\bullet \: \mathbf{2_{eve}}$]
   In this case
   {\small
   $$
      G = \pmatrix 0& iq\cr -iq^2& 0 
          \endpmatrix \sim \diag(1,-1) \ .
   $$}

\end{itemize}

%%%%%%%%%%%%%%%%%%%%%%%%%%%%%%%%%%%%%%%%%%%%%%%%%%%%%%%%%%%%%%%%%%%%%% 

\subsubsection{Invariant scalar products on $\mathcal M$} 
\label{s:inv_scalar_product_on_M}

It can be seen that for $N=3, q^3=1$ the reduced quantum plane 
$\mathcal M$, a module algebra for $\mathcal H$, is isomorphic as an 
algebra to the matrix algebra $M(3,\CC)$, whereas as a vector space 
splits into the sum of three {\it unequivalent} indecomposable 
representations, namely 
$\mathcal M \sim 3_{irr} \oplus 3_{eve} \oplus 3_{odd}$.

Actually, this feature can be generalized for all $N$ odd, $q^{N}=1$.
The corresponding quantum plane (which is now isomorphic with 
$M(N,\CC)$) splits into the sum of $N$ {\it unequivalent} 
indecomposable representations of $\mathcal H$. One of them is the 
irreducible $N_{irr}$, and the others are analogous to the 
``intermediate modules'' that appear within each lattice of 
submodules associated to the other $N-1$ projective indecomposable 
modules of $\mathcal H$. This property was proven in 
\cite{Coquereaux-Schieber}.

A word of warning seems here necessary: the algebra $M(N,\CC)$ 
plays an ubiquitous role here. Indeed, on one hand it is isomorphic 
with a simple subalgebra of $\mathcal H$ (see the structure of the 
regular representation given in \ref{a:structure_of_H}). As such, its 
underlying vector space splits into a sum of $N$ subspaces carrying 
equivalent representations (all equivalent to the $N_{irr}$), appearing 
in the decomposition of the regular representation in projective 
indecomposable modules. In this way $M(N,\CC)$ appears as an algebra 
and as a module, but not as a module algebra (considering 
$Z,W\in M(N,\CC) \subset \mathcal{H}$ and $X\in \mathcal{H}$, in general
$X(ZW) \neq (X_1 Z)(X_2 W)$). On the other hand, $M(N,\CC)$ is also 
isomorphic with the reduced quantum plane, and as such it is a module 
algebra, but not a subalgebra of $\mathcal H$ anymore. Its decomposition 
under the action of $\mathcal H$ is now more subtle, since it reads 
$M(N,\CC) \sim N_{irr} \oplus 
    N_{1} \oplus N_{2} \oplus \cdots \oplus N_{N-1}$. 

Because of this last result, one could be tempted to think that the 
most general scalar product on the reduced quantum plane $\mathcal M$ 
is simply given by the direct sum of its restrictions to the modules 
$3_{irr}$, $3_{eve}$ and $3_{odd}$ that are already known\ldots but 
it is not so. Indeed, non diagonal blocks may appear as we can have 
non-zero projections amongst vectors of different indecomposable 
representations.

Being $\mathcal M$ not only a module but a module algebra, we impose 
condition (\ref{star_representation}) for the left action of 
$\mathcal M$ on itself given by multiplication as well. This singles 
out a unique invariant hermitian form $(\:,\,)$, up to an overall 
scaling factor. Its structure was studied in Section~$5.5$ of 
\cite{CoGaTr-e} and goes as follows: the only non zero scalar products 
are those of the type $(x^r y^s,x^p y^t)$ with $r+p=s+t=2$, and they 
are all determined by setting $(xy,xy)=1$. The signature of this metric 
is $(5+,4-)$, so its index is $4$ and the Witt decomposition reads 
$9=4+4+1$. In the basis 
$\{\{ x^2,xy,y^2 \},\{ x,y,x^2 y^2 \},\{ \one,x^2 y,xy^2 \}\}$ 
the scalar product can be written as 
$$
G = \pmatrix B & 0 & 0 \cr
             0 & 0 & B \cr
             0 & B & 0 \endpmatrix \ ,
$$
where $B$ is the $3 \times 3$ block
$$
B = \pmatrix 0 & 0 & q^2 \cr
             0 & 1 & 0   \cr
             q & 0 & 0   \endpmatrix \ .
$$
The restriction of this scalar product to the subspace $3_{irr}$ 
coincides with what was already obtained before, a form of signature 
$(2+,1-)$. The restriction to the subspaces $3_{eve}$ and $3_{odd}$ is 
actually totally degenerate, so the conclusion we find for $\mathcal M$ 
does not contradict what was already obtained for $3_{eve}$ and 
$3_{odd}$ (just choose an overall scaling factor equal to $0$ in the 
latter cases).

%%%%%%%%%%%%%%%%%%%%%%%%%%%%%%%%%%%%%%%%%%%%%%%%%%%%%%%%%%%%%%%%%%%%%% 

\subsubsection{Invariant scalar products on the regular representation 
               of $\mathcal H$}
\label{s:scalar_product_on_H}

One should not be tempted to think that the most general hermitian 
scalar product on $\mathcal{H}$ $(N=3)$ itself is simply given by 
its restrictions to the direct sum 
$3 [3_{irr}]\oplus 2 [6_{eve}] \oplus 1 [6_{odd}]$ since we may 
very well accept ``off-block'' components. As a matter of fact, the 
constraints in this case are rather weak: 
for any given star, any hermitian form such 
that $(X^{*} Y, Z) = (Y, XZ)$ will work, but such a form is totally 
determined by the values of $(\one, X_{+}^a X_{-}^b K^c)$. Since we 
have $N^{3}$ terms, we see at once that the most general invariant 
scalar product on $\mathcal H$ will depend on $N^{3}$ parameters 
(real ones, due to the hermiticity of the scalar product). 
If one really wants to obtain an explicit expression for the possible 
metrics $G$'s, in the case $N=3$, the thing to do is to write 
explicitly $X_{\pm}$ and $K$ as $27 \times 27$ matrices (this is 
numerically easy, once we know how to write these generators in 
$M(3,\CC) \oplus \Mlo{2|1}$; this was done in \cite{Coquereaux} and 
recalled in \cite{CoGaTr-e}) and solve the equations
(\ref{star_representation_in_basis}), $||h||^\dag G = G ||h^*||$, 
for the coefficients $G_{ij}$ where $h=X_{\pm}$ and $K$.
One can then check that this set of equations indeed lead to a 
solution depending on $27$ parameters.

Because of this pretty big number of free parameters, the signature 
can be rather arbitrary. This is a slightly dispointing result since 
we are looking for some kind of constraint(s) that more or less fix 
the hermitian form. It would also be nice if the structure of this 
scalar product could somehow reflect the algebraic structure of 
$\mathcal H$ itself. As we shall see later, this goal will be 
achieved by the choice of a particular scalar product that we call 
the ``hermitian Killing form''. Yet another interesting scalar product
on the regular representation can be defined by using the existence
of left (or right) invariant integrals 
(see Section~\ref{s:scalar_products_with_integrals}).

%%%%%%%%%%%%%%%%%%%%%%%%%%%%%%%%%%%%%%%%%%%%%%%%%%%%%%%%%%%%%%%%%%%%%% 
%%%%%%%%%%%%%%%%%%%%%%%%%%%%%%%%%%%%%%%%%%%%%%%%%%%%%%%%%%%%%%%%%%%%%% 

\subsection{Invariant scalar products for $\mathcal H$ with a twisted 
            star}

\subsubsection{Invariant scalar products on the indecomposable 
               representations of $\mathcal H$}
\label{s:twisted_metrics_on_H_representations} 

As it was done for the true Hopf star in Appendix~E of 
\cite{CoGaTr-e}, we here show the most general metric on the vector 
space of each of the indecomposable representations of $\mathcal H$ 
using the twisted stars (\ref{twisted_star_on_H}). Since we have two 
possible choices, the $\pm$ signs below correspond respectively 
to the $\pm$ possibilities defined in (\ref{twisted_star_on_F}), 
(\ref{twisted_star_on_H}), (\ref{twisted_star_on_M}). We restrict 
the inner product to be a quantum group invariant one, as defined 
in Section~\ref{s:invariant_scalar_product_for_Hopf}. On each 
representation space we use the basis obtained from appropriate 
restrictions of the natural basis (``elementary basis'') associated 
with the regular representation of $\mathcal H$ as given in 
\ref{a:structure_of_H}. For each indecomposable representation we give 
an explicit expression of the most general covariant metric in this 
particular base and we calculate its signature.

\medskip

\begin{itemize}

\item[$\bullet \: \mathbf{3_{irr}}$]
Up to a real global normalization the metric is 
$$
G = \diag(1,\mp 1, 1) \ , \text{with signature}\ \sigma = (++\mp) \ .
$$

\medskip

\item[$\bullet \: \mathbf{6_{odd}}$]
Now we get the metric ($\beta \in \RR$)
{\small 
$$
G = \pmatrix 1&	0&	0& 0&	0& 0\cr
0&  \pm 1&	0& 0&	0& 0\cr
0&	0&  \pm 1& 0&	0& 0\cr
0&	0&	0& 1&	0& 0\cr
0&	0&	0& 0& \beta& 1\cr
0&	0&	0& 0&	1& 0 \endpmatrix \ .
$$}

\noindent
A change of basis tells us that
$G \sim \diag(1,1,\pm 1,\pm 1,\lambda_+,\lambda_-)$, with $\lambda_+ 
> 0, \lambda_- < 0$. Thus, the signature is $$
\sigma = (+++\pm \pm -) \ .
$$

\medskip

\item[$\bullet \: \mathbf{5_{odd}}$]
If $\alpha,\beta \in \RR$ and $g \in \CC$ we may write the metric as 
{\small 
$$
G = \pmatrix \alpha&	g&	0&	0& 0\cr
\bar g& \beta&	0&	0& 0\cr
0&	0& \pm\alpha& \pm g& 0\cr
0&	0& \pm\bar g& \pm\beta& 0\cr
0&	0&	0&	0& 0 \endpmatrix
$$}

\noindent
Given that
$G \sim \diag(\lambda_1,\lambda_2,\pm\lambda_1,\pm\lambda_2,0)$, with 
arbitrary $\lambda_i \in \RR$, its signature may be anything between 
$\sigma = (++++0)$ and $\sigma = (----0)$.

\medskip

\item[$\bullet \: \mathbf{3_{odd}}$]
Here
$$
G = \diag( 1, \pm 1, 0) \qquad \text{and} \ \sigma = (+\pm 0) \ .
$$

\medskip

\item[$\bullet \: \mathbf{6_{eve}}$]
Up to a normalization the metric can be written ($\beta \in \RR$) 
{\small 
$$
G = \pmatrix \beta& 1&	0&	0&	0& 0\cr
1& 0&	0&	0&	0& 0\cr
0& 0& \pm\beta& \pm 1&	0& 0\cr
0& 0& \pm 1&	0&	0& 0\cr
0& 0&	0&	0& \mp 1& 0\cr
0& 0&	0&	0&	0& -1 \endpmatrix
$$}

\noindent
The signature is clearly
$$
\sigma = (++\mp - - -) \ .
$$

\medskip

\item[$\bullet \: \mathbf{4_{eve}}$]
In this case the result coincides with the one obtained using the 
normal Hopf star, as we get the metric 
($\alpha, \beta \in \RR$, $g \in \CC$) 
{\small 
$$
G = \pmatrix 0& 0&	0&	0\cr
0& 0&	0&	0\cr
0& 0& \alpha&	g\cr
0& 0& \bar g& \beta \endpmatrix
$$}

\noindent
The non-null block in $G$ is an arbitrary hermitian matrix, 
therefore the signature is not fixed.

\medskip

\item[$\bullet \: \mathbf{3_{eve}}$]
As in the untwisted case, here we find simply 
$$
G = \diag( 0, 0, 1) \ . 
$$

\medskip

\item[$\bullet \: \mathbf{2_{eve}}$]
This irreducible representation has the metric 
$$
G = \diag( 1, \pm 1) \ .
$$
Notice that a positive definite form is obtained for the twisted Hopf 
star of $SU(2)$ type.

\end{itemize}

%%%%%%%%%%%%%%%%%%%%%%%%%%%%%%%%%%%%%%%%%%%%%%%%%%%%%%%%%%%%%%%%%%%%%% 

\subsubsection{Invariant scalar product on $\mathcal M$}

A priori, one could think that the discussion goes along the lines of 
Section~\ref{s:inv_scalar_product_on_M} and that nothing much should be 
changed. This is almost so, in the sense that invariance implies that 
the only {\it possibly non zero} scalar product of type $(\one,z)$ is 
$(\one,x^{2}y^{2})$. However, we shall show that this quantity vanishes 
as well (the proof uses the left action of $\mathcal{H}$ on 
$\mathcal{M}$ as discussed in Table~$1$, Section~$4.4$ of 
\cite{CoGaTr-e}). Indeed
\begin{eqnarray*}
(\one,x^{2}y^{2}) &=& (x^{2},y^{2}) = q^{-1}(x^{2},Ky^{2})
                      = q^{-1}(K^{*}x^{2},y^{2}) \\
    {} &=& q^{-1}(K^{-1}x^{2},y^{2}) = q^{-1}(qx^{2},y^{2}) \\
    {} &=& q^{2}q^{-1}(x^{2},y^{2}) = q(\one,x^{2}y^{2})
\end{eqnarray*}
Hence $(\one,x^{2}y^{2}) = 0$, and we see that the bilinear form obtained 
on $\mathcal{M}$ is totally degenerate. This result contrasts 
drastically with the one obtained in the untwisted Hopf star case.

\subsubsection{Invariant scalar products on the regular representation 
               of $\mathcal H$}

We refer to Section~\ref{s:scalar_product_on_H} for the general discussion 
and to the next section for a study of very specific scalar products on 
this representation space.

%%%%%%%%%%%%%%%%%%%%%%%%%%%%%%%%%%%%%%%%%%%%%%%%%%%%%%%%%%%%%%%%%%%%%%%%%% 
%%%%%%%%%%%%%%%%%%%%%%%%%%%%%%%%%%%%%%%%%%%%%%%%%%%%%%%%%%%%%%%%%%%%%%%%%% 

\section{Scalar products on the left regular representation of a Hopf 
         algebra}
\label{s:scalar_products_on_left_rr}

\subsection{The hermitianized Killing form} 
\label{s:hermitianized_Killing_form}

As is recalled in \ref{a:Killing_form}, in the case of Hopf algebras 
there is still a notion of a Killing form, which generalizes this 
particular bilinear form found in the case of Lie groups and algebras.
Moreover, it is also invariant under an adequate generalization of the 
adjoint action of a group on itself, now a left action of a Hopf algebra 
on itself.

This Killing form $(.\,,.)_{u}$ is neither symmetric nor hermitian 
(actually we did not use any star in its definition), but, 
given an arbitrary star operation on $H$, we now define
a sesquilinear form on $H \times H$ by\footnote{
As for the Killing form, there is always an implicit choice of 
representation of $H$, here the left regular one.
}
\begin{equation}
   (X,Y) \doteq (X^*,Y)_u = \tr_q(X^*Y) \ , \qquad X,Y \in H \ .
\end{equation}
This new form is obviously $H$-invariant ---in the sense of 
(\ref{star_representation})--- under the left action of $H$ on itself 
given by simple multiplication, as 
\begin{equation}
   (XY,Z) = \tr_q(Y^* X^* Z) = (Y,X^* Z) \ .
\label{multiplication_invariance_of_hermitianized_Killing}
\end{equation}

\noindent
The ``symmetry'' property of the Killing form (\ref{symmetry_of_Killing}) 
gets traduced now in
$$
   (Y,X) = (X^*, S^2(Y^*)) \ .
$$

\noindent
In addition, {\it if} the star operation is a true Hopf one, the 
invariance of $(\,,)_u$ under the adjoint action 
(\ref{invariance_of_Killing}) implies that
\begin{equation}
\left( ad_{(SZ_1)^*}(X), ad_{Z_2}(Y) \right) = 
    \left( X,Y \right) \, \epsilon (Z) \ .
\label{adjoint_invariance_of_hermitianized_Killing}
\end{equation}

\noindent
This is so because $[ad_{(SZ)^*}(X)]^* = ad_{Z}(X^*)$ for a Hopf star.
Note that both properties
(\ref{multiplication_invariance_of_hermitianized_Killing}) and 
(\ref{adjoint_invariance_of_hermitianized_Killing}) are invariances of 
this Killing scalar product in the sense of 
(\ref{invariant_scalar_product}), but with respect to different actions.
Actually, we also have for this action a $*$-representation, as it is 
true that
\begin{equation}
   \left( ad_Z(X), Y \right) = \left( X,ad_{Z^*}(Y) \right) \ . 
   \nonumber
\label{*-representation_of_hermitianized_Killing}
\end{equation}

Finally, when the star involved in this definition is a Hopf star, the 
resulting form is ---or can always be chosen to be--- hermitian; we call 
it the ``hermitianized Killing form'' or the ``Killing scalar product''
(we will see later that this is not the case when one uses a twisted star). 
Indeed, as we are working with a $*$-representation,
$\tr[h^*] = \overline{ \tr[h]} \,,\ h \in H$. Therefore, 
\begin{eqnarray*}
   (X,Y) &=& \tr(u\,X^* Y) = \overline{\tr(Y^* X u^*)} \\
         &=& \overline{\tr(u^* Y^* X)} \ ,
\end{eqnarray*}
and 
\begin{eqnarray}
   (X,Y) &=& \overline{(Y,X)}
\end{eqnarray}
if $u^* = u$. Using the notation of \ref{a:Killing_form}, we know that
$S^2(h) = u h u^{-1}$ implies, for a Hopf star, 
$S^2(h) = u^* h {(u^*)}^{-1}$. Both equations together tell us that
$u^{-1} u^*$ is a central element which, being a matrix on a 
representation space, should be proportional to the identity. Moreover 
the proportionality factor must be a phase ($(u^*)^* = u$), and this 
may always be absorbed in $u$ to have an hermitian form.

%%%%%%%%%%%%%%%%%%%%%%%%%%%%%%%%%%%%%%%%%%%%%%%%%%%%%%%%%%%%%%%%%%%%%%%%%% 

\subsubsection{The Killing scalar product for $\mathcal{H}$. Hopf star case.}
\label{s:Killing_scalar_product_on_H}

We have just defined a particular scalar product based on the Killing 
form on the regular representation of a quantum group $H$. We analyze 
here the case of the finite Hopf algebra $\mathcal H$, taking $N=3$, and
we choose a Hopf star operation. Then
$$
   (X,Y) = \tr_q(X^*Y) = \tr(K^{-1} X^*Y) \ , \qquad X,Y \in \mathcal{H} \ .
$$

In this case, the structure of the corresponding $27\times 27$ hermitian 
matrix $G$ in the PBW-basis is not very transparent and we shall not 
give it explicitly, although its signature can be read off easily. 
However, the expression of $G$ in what we called the ``elementary basis'' 
is quite remarkable. Here it goes: 

\begin{itemize}

\item Its restriction to the $M(3,\CC)$ block, with basis ordering 
$$
   \{E11,E12,E13,E21,E22,E23,E31,E32,E33\} \ ,
$$
reads:
$$
3 \pmatrix
0 & 0 & 0 & 0 & 0 & 0 & 0 & 0 & q^{-1}  \\
0 & 0 & 0 & 0 & 0 & 0 & 0 & -q^{-1} & 0 \\
0 & 0 & 0 & 0 & 0 & 0 & q^{-1} & 0 & 0  \\
0 & 0 & 0 & 0 & 0 & -1 & 0 & 0 & 0 \\
0 & 0 & 0 & 0 & 1 & 0 & 0 & 0 & 0  \\
0 & 0 & 0 & -1 & 0 & 0 & 0 & 0 & 0 \\
0 & 0 & q & 0 & 0 & 0 & 0 & 0 & 0  \\
0 & -q & 0 & 0 & 0 & 0 & 0 & 0 & 0 \\
q & 0 & 0 & 0 & 0 & 0 & 0 & 0 & 0
\endpmatrix \ .
$$

\item Its restriction to the subspace $\{A11,A12,A21,A22\}$ of the 
$\Mlo{2|1}$ block reads
$$
6 \pmatrix
0 & 0 & 0 & q       \\
0 & 0 & -q & 0      \\
0 & -q^{-1} & 0 & 0 \\
q^{-1} & 0 & 0 & 0
\endpmatrix \ .
$$

\item Its restriction to the subspace $\{A33\}$ of the $\Mlo{2|1}$ 
block reads
$$
6 \pmatrix 1 \endpmatrix \ .
$$

\end{itemize}

\noindent
All other entries vanish. This is in particular so for the scalar 
products mixing the three aforementioned subspaces. Also vanish
all scalar products between vectors belonging to the $13$ 
dimensional radical spanned by the generators 
$$
   \{B11,B12,B21,B22,P13,Q13,P23,Q23,P31,Q31,P32,Q32\} \ .
$$

In other words, $G$ is completely degenerated\footnote{
The fact that the trace of the adjoint map vanishes on the radical, 
a result slightly weaker that the one reported here, was separately 
observed by \cite{Kastler}.
}
in the direction of the Jacobson radical of $\mathcal H$, and it does 
not mix the different simple components of the semi-simple part 
$\overline{\mathcal H} \doteq M(3,\CC) \oplus M(2,\CC) \oplus \CC$. 
Moreover, we see at once that $G$, restricted to $\overline{\mathcal H}$ 
is diagonal in the (hence orthogonal) basis 
\begin{eqnarray*}
   & &\{ E11 \pm q^{-1} E33, E12 \pm q^{-1} E32, E13 \pm q^{-1} E31, 
         E21 \pm q^{-1} E23, E22 \,; \\
   & & \qquad A11 \pm q^{-1} A22, A12 \pm q^{-1} A21 \,;\, A33 \} \ ,
\end{eqnarray*}
where it actually reads:
$$ 
  G = 3 \, \diag \, (\pm 1,\pm 1, \pm 1, \pm 1, 1\,;
                     \, \pm 2, \pm 2 \,;\, 1) \ .
$$

\noindent
The signature of the restriction of $G$ to $\overline{\mathcal H}$ 
reads therefore $(8+,6-)$, but it is better to write it (with obvious 
notations) as
$$
  \left[4 (+1,-1) \oplus    (+1) \right] \oplus 
  \left[  (+1,-1) \oplus (+1,-1) \right] \oplus (+1) \ .
$$ 

%%%%%%%%%%%%%%%%%%%%%%%%%%%%%%%%%%%%%%%%%%%%%%%%%%%%%%%%%%%%%%%%%%%%%%%%%% 
\subsubsection{Incompatibility between a Killing scalar product 
               and a twisted Hopf star}

Here we can follow a discussion along the same lines of the last part of 
Section~\ref{s:hermitianized_Killing_form}, but now starting from 
$S^2(h) = u h u^{-1}$ it is easy to deduce that 
$S^2(h) = {(u^*)}^{-1} h u^*$. Both formulas together imply that 
$u u^*$ is a central element, and this means that we will have 
$u^* = cu^{-1} \neq u$ ($c \in H$ central).

Therefore, we can not expect to have an hermitian Killing scalar product
if the star is a twisted one. Having a true (hermitian) scalar product 
is incompatible with the invariance of the Killing form.

%%%%%%%%%%%%%%%%%%%%%%%%%%%%%%%%%%%%%%%%%%%%%%%%%%%%%%%%%%%%%%%%%%%%%%%%%% 
%%%%%%%%%%%%%%%%%%%%%%%%%%%%%%%%%%%%%%%%%%%%%%%%%%%%%%%%%%%%%%%%%%%%%%%%%% 

\subsection{Scalar products related to invariant integrals} 
\label{s:scalar_products_with_integrals}

We first gather general facts and definitions about left and right 
invariant integrals on a Hopf algebra. We then use these
concepts ---together with a star operation--- to define a 
particular hermitian scalar product on finite dimensional Hopf
algebras. All these notions are illustrated with our favorite 
example $\mathcal H$.

\subsubsection{Integrals}

A left-invariant integral on a Hopf algebra $H$ over $\CC$ is a 
linear map $\int_{L} : H \mapsto \CC$ such that
$$
   \left( \id \otimes \int_{L}\right)\circ \Delta = 
   \one_{H} \int_{L} \ ,
$$
where $\one_{H}$ is the unit of $H$ and $\id$ the identity map in $H$.
Therefore, for any $h \in H$ we have (as always 
$\Delta h = h_1 \otimes h_2$)
\begin{equation}
   h_1 \int_{L} h_2 = \one_{H} \int_{L} h \ .
\label{L_invariance_of_integral}
\end{equation}

\noindent
A right-invariant integral $\int_R$ is defined in the obvious similar 
way. 

Since $\int_{L}$ (or $\int_{R}$) is a linear object, it can be 
identified with an element $\lambda_{L}$ (resp. $\lambda_{R}$) of the 
dual $F$ of $H$. Such an element will therefore satisfy
$$
f \lambda_{L} = \epsilon(f) \lambda_{L}
$$
(or $\lambda_{R} f = \epsilon(f) \lambda_{R}$) for any $f \in F$.

Like for groups, a Hopf algebra $H$ is called {\it unimodular} if one 
can find left and right integrals which coincide 
($\int \doteq \int_{L} = \int_{R}$). Furthermore, such an integral 
is called a Haar measure when it is normalizable and normalized, \ie 
$\int(\one_{H}) = 1$ (in particular $\int$ should not vanish on the 
unit!). 

We now go back to the example where $H$ is a reduced quantum
enveloping algebra of type $SL_{q}(2,\CC)$ at a root of unity, 
$\mathcal H$. It is easy to see that here the left and right 
integrals are respectively given (up to an overall constant) by
$$
\int_{L} = (X_{+}^{N-1} X_{-}^{N-1} K)^\star
$$
and
$$
\int_{R} = (X_{+}^{N-1} X_{-}^{N-1} K^{-1})^\star \ .
$$

\noindent
Here a particular vector space basis (PBW) 
$\{X_{+}^{a}X_{-}^{b}K^{c}\}$ is chosen in $\mathcal H$ and
$\{(X_{+}^{a}X_{-}^{b}K^{c})^\star\}$ denotes its dual basis.
In terms of elements of $\mathcal F$, the same left and right 
invariant integrals on $\mathcal H$ read
\begin{eqnarray*}
\lambda_{L} &=& (1 + a + \ldots + a^{N-1}) b^{N-1} c^{N-1} \\
\lambda_{R} &=& b^{N-1} c^{N-1} (1 + a + \ldots + a^{N-1}) \ .
\end{eqnarray*}

\noindent
These two integrals are not proportional and cannot be made equal; 
$\mathcal H$ is therefore not unimodular and no Haar measure can be 
defined. The dual $\mathcal F$ of $\mathcal H$ turns out to be 
unimodular (see \cite{Dabrowski}), but the corresponding integral is 
not a Haar measure because it is not normalizable as it vanishes on 
the unit. 

Further restricting now our class of examples to the case $N = 3$, it 
is interesting to decompose the elements $X_+^2 X_-^2 K$ and 
$X_+^2 X_-^2 K^{-1}$ on the elementary basis defined in the 
\ref{a:structure_of_H}. They read respectively:
{\small
$$
\pmatrix
\pmatrix
q^{2} & 0 & 0 \cr
0 & 0 & 0 \cr
0 & 0 & 0
\endpmatrix & 0 \cr
0 & \pmatrix
-q \theta^{1}\theta^{2} & 0 & 0 \cr
0 & 0 & 0 \cr
0 & 0 & \theta^{1}\theta^{2}
\endpmatrix
\endpmatrix
$$
}
and
{\small
$$
\pmatrix
\pmatrix
q & 0 & 0 \cr
0 & 0 & 0 \cr
0 & 0 & 0
\endpmatrix & 0 \cr
0 & \pmatrix
-q^{2} \theta^{1}\theta^{2} & 0 & 0 \cr
0 & 0 & 0 \cr
0 & 0 & \theta^{1}\theta^{2}
\endpmatrix
\endpmatrix \ .
$$
}

\noindent On the other hand, using the PBW basis, the invariant 
integral\footnote{
Notice that on the group $\ZZ_{3} = \{\one, K, K^2\}$
the integral is given by $\Sigma = 1 + K + K^{2}$.
}
on $\mathcal F$ can be expressed by duality as the element 
$\sigma \doteq X_+^{2}X_-^{2} (1 + K + K^{2}) \in \mathcal{H}$. 

%%%%%%%%%%%%%%%%%%%%%%%%%%%%%%%%%%%%%%%%%%%%%%%%%%%%%%%%%%%%%%%%%%%%%%%%%% 

\subsubsection{Scalar product on the left regular representation} 

Using both a star operation (any) and an integral on $H$, we now 
define a kind of Hopf algebra analogue of the familiar scalar product 
used to discuss square integrable functions in usual complex 
analysis. We take
\begin{equation}
   (X,Y)_{L,R} \doteq \int_{L,R} X^{*} Y \ ,
\end{equation}
which is then automatically sesquilinear and invariant. In fact, by 
construction this scalar product satisfies the $*$-representation 
condition, as
$$
(ZX,Y) = (X,Z^* Y) \ .
$$
Here $H$ acts on itself by left-multiplication, and the invariance is 
independent of the star chosen (twisted or not). 

Other properties of this scalar product will of course depend upon 
the kind of star used in its definition.

\paragraph{The Hopf star case}

\begin{itemize}

\item
To have hermiticity of our scalar product we need only to check that 
$$
   \int_{L,R} X^* = \overline{\int_{L,R} X} \ ,
$$
as $(Y,X) = \int{(X^*Y)^*}$ and 
$\overline{\int{X^*Y}}=\overline{(X,Y)}$.
It is easy to see that the above property is compatible with the 
left-invariance of this integral (contrarily to what will happen in 
the twisted star case). Therefore one needs to check this explicitly 
for each case, knowing that a left (or right) invariant integral on a 
Hopf algebra is unique ---if it exists---, up to a scalar multiple. 
We checked explicitly this property for the case of $H=\mathcal{H}$.

\item
{}From the invariance property of $\int_{L}$, one trivially gets 
$$
\one_H \, (X,Y) = X_1^* Y_1 \, (X_2,Y_2) \ .
$$
But this may also be interpreted ---as happens with the integral--- 
as an invariance with respect to the right action of $F$: 
\begin{eqnarray*}
(X,Y)\triangleleft f &=& \epsilon(f) \, (X,Y) \\ 
         &=& (X \triangleleft (Sf_1)^*\,,\, Y \triangleleft f_2) \ . 
\end{eqnarray*}
This expression is the analogue of (\ref{invariant_scalar_product_2}) 
for a right action.

Recall that this is an extra invariance of the scalar product, as by 
construction it is invariant under the left action of $H$ itself.

\item
In our example of $\mathcal H$, with $N=3$, this hermitian form 
expressed in terms of the ``elementary basis'' defined in 
\ref{a:structure_of_H} gives a $27 \times 27$ hermitian matrix 
$G_{ij}$ that we describe now. Its restriction to the $9$-dimensional 
subspace spanned by
$$
\{E_{11},E_{12},E_{13},E_{21},E_{22},E_{23},E_{31},E_{32},E_{33}\}
$$
reads
{\small
$$
\frac{1}{3} \pmatrix
0 & 0 & 0 & 0 & 0 & 0 & 0 & 0 &q^{-1}\cr
0 & 0 & 0 & 0 & 0 & 0 & 0 &-q^{-1}& 0 \cr
0 & 0 & 0 & 0 & 0 & 0 &q^{-1} & 0 & 0 \cr
0 & 0 & 0 & 0 & 0 &-1 & 0 & 0 & 0 \cr
0 & 0 & 0 & 0 &1 & 0 & 0 & 0 & 0 \cr
0 & 0 & 0 &-1 & 0 & 0 & 0 & 0 & 0 \cr
0 & 0 &q & 0 & 0 & 0 & 0 & 0 & 0 \cr
0 &-q & 0 & 0 & 0 & 0 & 0 & 0 & 0 \cr
q & 0 & 0 & 0 & 0 & 0 & 0 & 0 & 0 \cr
\endpmatrix \ .
$$
}

\noindent
Its restriction to the $2(4)+2(1)=10$-dimensional subspace spanned by
$$
\{A_{11},B_{11},A_{12},B_{12},A_{21},B_{21},A_{22},B_{22},A_{33},B_{33}\}
$$
reads
{\small
$$
\frac{1}{3} \pmatrix
0 & 0 & 0 & 0 & 0 & 0 &-q &-q & 0 & 0 \cr
0 & 0 & 0 & 0 & 0 & 0 &-q & 0 & 0 & 0 \cr
0 & 0 & 0 & 0 &q &q & 0 & 0 & 0 & 0 \cr
0 & 0 & 0 & 0 &q & 0 & 0 & 0 & 0 & 0 \cr
0 & 0 &q^{-1} &q^{-1} & 0 & 0 & 0 & 0 & 0 & 0 \cr
0 & 0 &q^{-1} & 0 & 0 & 0 & 0 & 0 & 0 & 0 \cr
q^{-1} &-q^{-1} & 0 & 0 & 0 & 0 & 0 & 0 & 0 & 0 \cr
q^{-1} & 0 & 0 & 0 & 0 & 0 & 0 & 0 & 0 & 0 \cr
0 & 0 & 0 & 0 & 0 & 0 & 0 & 0 & -1 & 1 \cr
0 & 0 & 0 & 0 & 0 & 0 & 0 & 0 & 1 & 0 \cr
\endpmatrix \ .
$$
}

\noindent
Finally, its restriction to the $8$-dimensional subspace spanned by
$$
\{P_{13},Q_{13},P_{23},Q_{23},P_{31},Q_{31},P_{32},Q_{32}\}
$$
reads
{\small
$$
\frac{1}{3} \pmatrix
0 & 0 & 0 &q & 0 & 0 & 0 & 0 \cr
0 & 0 &-q & 0 & 0 & 0 & 0 & 0 \cr
0 &-q^{2} & 0 & 0 & 0 & 0 & 0 & 0 \cr
q^{2} & 0 & 0 & 0 & 0 & 0 & 0 & 0 \cr
0 & 0 & 0 & 0 & 0 & 0 & 0 & 1 \cr
0 & 0 & 0 & 0 & 0 & 0 &-1 & 0 \cr
0 & 0 & 0 & 0 & 0 &-1 & 0 & 0 \cr
0 & 0 & 0 & 0 &1 & 0 & 0 & 0 \cr
\endpmatrix \ .
$$
}
All the other scalar products vanish.

First of all, we may notice at once that this hermitian form is not
degenerate (this sharply contrasts with the hermitianized Killing form 
which is degenerate along the radical, as we saw previously).
Here, the signature is $(14+,13-)$. The $27$ eigenvalues themselves read:
$$
\frac{1}{3} \left\{ (1)_9, (-1)_8, (\beta)_2 , 
           (-\beta^{-1})_2, (-\beta)_3, (\beta^{-1})_3 \right\} \ ,
$$
where $\beta = \frac{1 + \sqrt{5}}{2}$ is the golden number.
It is interesting to notice that, although non degenerate, the 
restriction of this form to the $9+4+1=14$-dimensional semi-simple part 
of $\mathcal H$ is positive definite (this part, isomorphic with the 
matrix algebra $M(3,\CC)\oplus M(2,\CC) \oplus \CC$, as recalled in 
\ref{a:structure_of_H}, is spanned by $E_{ij}$ and $A_{kl}$).

\end{itemize}

\paragraph{The twisted star case}

\begin{itemize}

\item
It is in general {\it not} hermitian. In fact, if we now write down 
(\ref{L_invariance_of_integral}) for $h^*$ and conjugate that 
equation, we get
$$
h_2 \, \overline{\int_L h_1^*} = \one \, \overline{\int_L h^*} \ .
$$
If we assume that $\int_L h^* = \overline{\int_L h}$, the above 
equation would tell us that $\int_L$ should also satisfy the {\it 
right} invariance condition, which will not be generally true. For 
instance, in the case of $\mathcal H$ we know that a biinvariant 
integral does not exist. To obtain an hermitian scalar product we 
could then add both integrals, $(X,Y) \doteq (\int_L + \int_R) X^* Y$, 
but this one would not have any extra invariance property\ldots

\item
{}From the invariance property of $\int_{L}$ results 
$$
\one_H \, (X,Y) = X_2^* Y_1 \, (X_1,Y_2) \ ,
$$
which shows a left-right mixed behaviour. 

\item
The example of $\mathcal H$, with $N=3$, is not particularly enlighting 
since the obtained complex bilinear form is not hermitian but symmetric. 
A numerical study of this $27 \times 27$ matrix in the elementary 
basis defined in \ref{a:structure_of_H} shows that it is not degenerate 
and that it is ``almost'' diagonal, in the sense that the only non 
diagonal $G_{ij}$ entries are $G(A_{11},B_{11})$, $G(A_{12},B_{12})$, 
$G(A_{21},B_{21})$, $G(A_{22},B_{22})$ and $G(A_{33},B_{33})$ together 
with the corresponding symmetric coefficients. We however stress 
again the fact that, using the twisted star, the scalar product is 
not hermitian.

\end{itemize}

To conclude: the twisted Hopf star case is rather bad in this sense.

%%%%%%%%%%%%%%%%%%%%%%%%%%%%%%%%%%%%%%%%%%%%%%%%%%%%%%%%%%%%%%%%%%%%%%%%%%
%%%%%%%%%%%%%%%%%%%%%%%%%%%%%%%%%%%%%%%%%%%%%%%%%%%%%%%%%%%%%%%%%%%%%%%%%%

\section{Discussion}
\label{s:discussion}

As it was mentioned in the Introduction, the parameter $q$ that appears
in many integrable and conformal models is often a primitive root of
unity, and such values are generally incompatible with the choice of a
compact real form on the quantum group (like $SU_q(2)$, for instance).
For this reason the stars on ``compact'' quantum groups that one may
define in the context of spin chains, for example, are twisted. The 
discussion is however a bit subtle and we want to make the following 
comments:

In the case of a spin chain of type $XXZ$, for instance (see 
\cite{Pasquier-Saleur}, for example), one may start with the 
{\sl usual} rotation group in three dimensions ---or with its double 
cover $SU(2)$--- acting at each point of the chain. Another ingredient 
is given by the choice of some (unitary) representation of this group, 
for instance the fundamental ($s=1/2$). The Hilbert space of the model 
is obtained as the $n$-th tensor product of this representation. The 
hamiltonian of the model is given by a sum of interaction terms indexed
by a discrete label, each term being itself built in terms of 
(hermitian) Pauli matrices. This hamiltonian is {\sl not}, in general, 
invariant with respect to the rotation group since the physical system 
is clearly not rotationally invariant. However, in some cases, one 
notices that the same total hamiltonian commutes with the generators 
of a (complex) quantum group, for instance $U_q(sl(2,\CC))$. We should 
stress the fact that generators of $SU(2)$ act on the Hilbert space in 
a way that is ``local'' (generators rotate the states independently at 
each point of the chain), whereas $U_q(sl(2,\CC))$ acts in a non local 
way (this point of view was emphasized for instance in 
\cite{Bernard-Leclerc}). Notice that hermiticity of the hamiltonian 
---a Jones projector--- is clearly a required constraint, however this 
property does not take place in a representation space for the quantum 
group but in its commutant.

Both $SU(2)$ and $U_q(sl(2,\CC))$ enter the discussion of the model and
both have two-dimensional representations, but the two related concepts 
should not be confused. For physical reasons, it is clear that the 
scalar product used on the Hilbert space of the model should not contain 
vectors of negative norm; for this reason it should be a {\it bona-fide} 
positive definite scalar product. The same Hilbert space could also be 
built in terms of tensor products of the fundamental representation of 
the quantum group $U_q(sl(2,\CC))$, for $q$ a root of unity; indeed, two 
vector spaces over $\CC$ of the same dimension are clearly isomorphic, as 
vector spaces\ldots Nevertheless, in the usual construction the Hilbert 
space of the model acquires its Hilbert structure from the scalar product 
chosen on representations of $SU(2)$, not from the one chosen on the
representations of the quantum group. Actually, the authors of the 
present paper do not see why such a choice should be performed at all; 
they cannot exclude however that it may turn out to be be useful. What 
is in any case clear, is that {\sl if} one wants to choose a scalar 
product on the fundamental representation of $U_q(sl(2,\CC))$ such that 
it will induce the same (already given) positive scalar product on the 
Hilbert space of the model, one has to suppose that the quantum group 
is endowed with a star operation which is a twisted Hopf star of 
$SU(2)$ type.

We should mention the papers \cite{Mack-Schomerus-1, Mack-Schomerus-2}, 
where a general study of quantum symmetries in quantum theory is done, 
and where the {\sl choice} of twisted star operations is clearly made 
right at the beginning. This was actually nothing else than a choice 
(related to a way of defining a covariant adjoint for field operators), 
and it was subsequently discovered\footnote{
Unpublished addendum by the same authors. We thank G. Mack for this
information.
}
that this choice was not unique and that it would have been also 
perfectly possible to define adjoints for field operators after having
decided to use a ``true'' Hopf star operation.

In conformal theories, primary fields are associated with vectors of 
highest weight in a representation of some affine algebra, and it was 
observed long ago that the fusion table of such primary fields is 
identical to the Clebsh Gordan table describing the tensor products of 
irreducible representations of some quantum group ---the same quantum 
group also appears, via its $6j$-symbols, in the equations describing 
the duality properties of the conformal blocks. At this point, one 
should stress that the representations of the quantum group that appear 
in the associated fusion table are not to be confused with the 
representations of the affine algebra. The two structures, although 
related (in a way that is apparently not well understood yet, see 
\cite{Alekseev-Faddeev}), are quite distinct and the discussion
involving the nature of the scalar product to be used in a given
representation space for the affine or Virasoro generators should 
not be confused with the analysis of the scalar product(s) that one 
can define on the modules of the emerging quantum group.

When the parameter $q$ is a root of unity, the representation theory
is quite subtle since indecomposable (but not irreducible)
representations of the quantum group appear. Actually, to obtain a
physically meaningful state space one has to choose a so-called
``truncated tensor product'', by selecting only those representations
for which the $q$-trace vanishes (one can also use the formalism of
quasi-Hopf algebras, see \cite{Mack-Schomerus-2}). It is a fact that
discussions involving quantum groups in conformal field theories usually
consider {\sl infinite dimensional} Hopf algebras (like $U_q(sl(2,\CC))$),
which are not ``good'' quantum groups when $q$ is a root of unity since
they are not quasi-triangular in the usual sense. At the contrary, the
{\sl finite dimensional} Hopf algebras that one can obtain from those
ones through division by an (infinite dimensional) Hopf ideal are not
semi-simple but they {\sl are} quasi-triangular: they possess (finite
dimensional) $R$-matrices. The category of representations of these
Hopf algebras is not a modular category (tensor products of irreducible
representations are not necessarily equivalent to direct sums of
irreducibles), but it is again possible to define truncated scalar
products in a very natural way. We conjecture that discussions involving
simultaneously rational conformal field theories and quantum groups
should be done in terms of such finite dimensional Hopf quotients of
the usual quantum enveloping algebras at roots of unity. A general
study of these topics stays outside the scope of the present paper
but we hope that our contribution concerning stars (twisted or not)
and scalar products, together with selected examples involving finite
dimensional Hopf algebra quotients of $U_q(sl(2,\CC))$ will be useful
in this respect.

%%%%%%%%%%%%%%%%%%%%%%%%%%%%%%%%%%%%%%%%%%%%%%%%%%%%%%%%%%%%%%%%%%%%%%%%%%
%%%%%%%%%%%%%%%%%%%%%%%%%%%%%%%%%%%%%%%%%%%%%%%%%%%%%%%%%%%%%%%%%%%%%%%%%%

\section*{Acknowledgements}

A. G. wishes to thank the CPT for the warm hospitality.
R. T. thanks CONICET, ICTP and the Univ. d'Aix Marseille I for financial
support. Also the kind hospitality extended to him at the CPT is 
gratefully acknowledged.

%%%%%%%%%%%%%%%%%%%%%%%%%%%%%%%%%%%%%%%%%%%%%%%%%%%%%%%%%%%%%%%%%%%%%%%%%% 
%%%%%%%%%%%%%%%%%%%%%%%%%%%%%%%%%%%%%%%%%%%%%%%%%%%%%%%%%%%%%%%%%%%%%%%%%% 

\appendix

\section*{Appendices}

\setcounter{section}{0}
\setcounter{subsection}{0}
\setcounter{subsubsection}{0}
\def\thesection{}
\def\thesubsection{Appendix \Alph{subsection}} 
\def\thesubsubsection{Appendix 
\Alph{subsection}.\Roman{subsubsection}} 

%%%%%%%%%%%%%%%%%%%%%%%%%%%%%%%%%%%%%%%%%%%%%%%%%%%%%%%%%%%%%%%%%%%%%%%%%% 

\subsection{Structure of the reduced Hopf algebra ${\mathcal H}$} 
\label{a:structure_of_H}

When $q$ is a root of unity ($q^{N}=1$), the quantized enveloping 
algebra $U_q(sl(2,\CC))$ possesses interesting quotients that are 
finite dimensional Hopf algebras. The structure of the left regular 
representation of such an algebra was investigated in \cite{Alekseev} 
and the pairing with its dual in \cite{Gluschenkov}.
We call ${\mathcal H}$ the Hopf algebra quotient of $U_q(sl(2,\CC))$
defined by the relations 
$$
K^{N}=\one \ , \qquad X_{\pm}^{N}=0 \ ,
$$
and ${\mathcal F}$ its dual.
The generators $K, X_{\pm}$ are chosen to obey the following 
commutation and cocommutation relations:
\begin{description}
\item[Product:]
\begin{eqnarray}
K X_{\pm} &=& q^{\pm 2} X_{\pm} K  \nonumber \\
\left[ X_+ , X_- \right] &=& \frac{1}{(q - q^{-1})} (K - K^{-1})
   \label{H-products} \\
K^N	&=& \one	\nonumber \\
X_+^N = X_-^N &=& 0	\nonumber
\end{eqnarray}

\item[Coproduct:]
\begin{eqnarray}
\Delta X_+ & = & X_+ \otimes \one + K \otimes X_+ \nonumber \\ 
\Delta X_- & = & X_- \otimes K^{-1} + \one \otimes X_- 
   \label{H-coproducts} \\
\Delta K & = & K \otimes K	\nonumber \\
\Delta K^{-1} & = & K^{-1} \otimes K^{-1} \nonumber
\end{eqnarray}
\end{description}

It was shown\footnote{
Warning: the authors of \cite{Alekseev} actually consider a Hopf 
algebra quotient defined by $K^{2N} = \one, X_{\pm}^{N}=0$, so their 
algebra is, in a sense, twice bigger than ours (see 
\ref{a:algebras_related_to_H}).
}
in \cite{Alekseev} that the {\underline{non}} semi-simple algebra 
${\mathcal H}$ is isomorphic with the direct sum of a complex matrix 
algebra and of several copies of suitably defined matrix algebras 
with coefficients in the ring $Gr(2)$ of Grassmann numbers with two 
generators. The explicit structure of those algebras (for any $N$), 
including the expression of generators themselves, was obtained by 
\cite{Ogievetsky}. Using these results, the representation theory 
of ${\mathcal H}$ for the case $N=3$ was presented in 
\cite{Coquereaux}. 

When $q^N=1$ with $N$ odd\footnote{
When $N$ is even with $N^\prime=N/2$ odd, $K^{N^\prime}$, 
$X_\pm^{N^\prime}$ are central and one may take the quotient by
$K^{N^\prime}=\one$, $X_\pm^{N^\prime}=0$; the algebra so obtained is 
isomorphic with $\mathcal{H}$. When $N^\prime=N/2$ is even the structure 
is quite different, and we do not study it here (see \cite{Ogievetsky}).
},
we have an isomorphism between the $N^{3}$-dimensional algebra 
${\mathcal H}$ and the direct sum
\begin{equation}
   \mathcal{H} = M_N \oplus \Mlo{N-1|1} \oplus \Mlo{N-2|2} \oplus \cdots 
   \oplus \Mlo{\frac{N+1}{2}|\frac{N-1}{2}}
\label{H_as_matrices_isomorphism} 
\end{equation}
where:
\begin{itemize}

\item[-] $M_N$ is a $N\times N$ complex matrix
\item[-] an element of the $M_{N-2|2}$ block (for instance) is of the kind: 
{\small
\begin{equation}
\begin{pmatrix}
\bullet & \bullet & \cdots & \bullet & \circ & \circ \cr
\bullet & \bullet & \cdots & \bullet & \circ & \circ \cr 
\vdots  & \vdots  &	   & \vdots  &\vdots & \vdots\cr
\bullet & \bullet & \cdots & \bullet & \circ & \circ \cr
\circ   & \circ   & \cdots & \circ   & \bullet & \bullet \cr
\circ   & \circ   & \cdots & \circ   & \bullet & \bullet \cr
\end{pmatrix}
\end{equation}
}
where we have introduced the following notation: \\ 
$\bullet$ is an even element of the ring $Gr(2)$ of Grassmann numbers 
with two generators\footnote{
Remember that $\theta_1^{2} = \theta_2^{2} = 0$ and 
$\theta_1 \theta_2 = -\theta_2 \theta_1$.
}, 
\ie of the kind:
$$
\bullet = \alpha + \beta \theta_1 \theta_2 \ , 
           \qquad \alpha,\beta \in \CC
$$
$\circ$ is an odd element of the ring $Gr(2)$, \ie of the kind:
$$
\circ = \gamma\theta_1 + \delta \theta_2 \ , 
           \qquad \gamma, \delta \in \CC
$$
\item[-] \etc
\end{itemize}

\noindent
Notice that ${\mathcal H}$ is \underline{not} a semi-simple algebra: 
its Jacobson radical ${\mathcal J}$ is obtained by selecting in 
equation (\ref{H_as_matrices_isomorphism}) the matrices with elements 
proportional to Grassmann variables. The quotient 
${\mathcal H} / {\mathcal J}$ is then semi-simple\ldots but no 
longer Hopf!

Projective indecomposable modules (PIM's, also called principal 
modules) for ${\mathcal H}$ are directly given by the columns of the 
previous matrices. 
\begin{itemize}
    \item[-] From the $M_{N}$ block, one obtains $N$ equivalent 
             irreducible representations of dimension $N$ that we shall 
             denote $N_{irr}$. These representations have vanishing 
             $q$-dimension.
    \item[-] From the $M_{{N-p}\vert p}$ block ($p<N-p$), one obtains
        \begin{itemize}
        \item $(N-p)$ equivalent indecomposable projective modules of 
              dimension $2N$ that we shall denote $P_{N-p}$ with 
              elements of the kind
              \begin{equation}
                 (\underbrace{\bullet \bullet \cdots \bullet}_{N-p} 
                 \underbrace{\circ \circ \cdots \circ}_{p}) \ .
              \end{equation} 

        \item $p$ equivalent indecomposable projective modules (also of
              dimension $2N$) that we shall denote $P_{p}$ with 
              elements of the kind 
              \begin{equation}
                 (\underbrace{\circ \circ \cdots \circ}_{N-p} 
                 \underbrace{\bullet \bullet \cdots \bullet}_{p}) \ .
              \end{equation}
        \end{itemize}
        These PIM's have also $q$-dimension equal to zero. To each PIM 
        $P_s$ is associated an irreducible representation of dimension 
        $s$, obtained by quotienting $P_s$ by its own radical. These  
        irreps have non vanishing $q$-dimension, and are in one to one 
        correspondence with the so called type II irreducible 
        representations of $U_q(sl(2,\CC))$.

\end{itemize}

\noindent
Other submodules can be found by restricting the range of parameters 
appearing in the columns defining the PIM's and imposing stability 
under multiplication by elements of ${\mathcal H}$. In this way one 
can determine for each PIM the lattice of its submodules. For each PIM 
of dimension $2N$, one finds totally ordered sublattices
with exactly three non trivial terms: the radical (here, it is the 
biggest non trivial submodule of a given PIM), the socle
(here it is the smallest non trivial submodule), and one 
``intermediate'' submodule of dimension exactly equal to $N$. However
the definition of this last submodule (up to an equivalence) depends on 
the choice of an arbitrary complex parameter
$\lambda$, so that we have a chain of inclusions for every such 
parameter. The collection of all these sublattices fully determines 
the lattice structure of submodules of a given principal module. 

We are interested in this paper in Hopf stars (twisted or not) and 
invariant scalar products for representation spaces of $\mathcal H$. 
To ease the presentation of the results, it is better to limit 
ourselves to the case $N=3$ but the overall picture should be clear. 
{}From now on, we take $N=3$.

In the case $q^{3}=1$, $\mathcal H$ is a $27$-dimensional Hopf algebra 
isomorphic with $M(3,\CC) \oplus \Mlo{2|1}$. Explicitly,
{\small
\begin{equation}
\mathcal H = \left\{ \pmatrix e_{11} & e_{12} & e_{13} \\
                      e_{21} & e_{22} & e_{23} \\
                      e_{31} & e_{32} & e_{33} \endpmatrix
\oplus
\pmatrix \alpha_{11} + \beta_{11} \theta_1 \theta_2 & \alpha_{12} + 
\beta_{12} \theta_1 \theta_2 & \gamma_{13} \theta_1 + \delta_{13} 
\theta_2 \\ \alpha_{21} + \beta_{21} \theta_1 \theta_2 & \alpha_{22} 
+ \beta_{22} \theta_1 \theta_2 & \gamma_{23} \theta_1 + \delta_{23} 
\theta_2 \\ \gamma_{31} \theta_1 + \delta_{31} \theta_2 & \gamma_{32} 
\theta_1 + \delta_{32} \theta_2 & \alpha_{33} + \beta_{33} \theta_1 
\theta_2 \endpmatrix \right\}\ .
\label{structure_of_H_matrices}
\end{equation}
}

\noindent All entries besides the $\theta$'s are complex numbers (the 
above $\oplus$ sign is a direct sum sign: these matrices are $6\times 
6$ matrices written as a direct sum of two blocks of size $3\times 
3$).

The semisimple part $\overline{\mathcal H}$, given by the direct sum 
of its block-diagonal $\theta$-independent parts is equal to the 
$9+4+1 = 14$-dimensional algebra $\overline{\mathcal H} = M_3(\CC) 
\oplus M_2(\CC) \oplus \CC$. The radical (more precisely the Jacobson 
radical) $J$ of $\mathcal H$ is the left-over piece that contains all 
the Grassmann entries, and only the Grassmann entries, so 
$\overline{\mathcal H} ={\mathcal H}/J$. The radical has therefore 
dimension $13$.

Projective indecomposable modules (PIM's) are given by the columns of 
the previous expression. We see that the left regular representation 
splits into a sum of three equivalent $3$-dimensional projective 
indecomposable representations that we call $3_{irr}$ (they are also 
irreducible) given by the columns of $M(3,\CC)$, two equivalent 
$6$-dimensional projective indecomposable representations that we call 
$6_{eve}$ given by the first two columns of $(M_{2\vert 1}(\Lambda^2))_0$ 
and one $6$-dimensional projective
indecomposable representation that we call $6_{odd}$ given by the 
last column of $(M_{2\vert 1}(\Lambda^2))_0$. The left regular 
representation can therefore be decomposed as follows:
$$
   3 [3_{irr}] \oplus 2 [6_{eve}] \oplus 1 [6_{odd}]
$$
All these projective indecomposable representations have zero 
quantum dimension.

Irreducible representations are obtained by taking the quotient of 
the projective indecomposable ones by their respective radical 
(killing the Grassmann variables). One obtains in this way the 
irreducible representation $3_{irr}$ that we already had, a two 
dimensional irreducible $2_{irr}$ (quotient of $6_{eve}$) and a one 
dimensional irreducible $1_{irr}$ (quotient of $6_{odd}$). Notice 
that $2_{irr}$ and $1_{irr}$ do not have vanishing quantum dimension, 
whereas as we already mentioned the $3_{irr}$ is special in this respect, 
since it is also one of the PIM's.

In order to discuss the results it is convenient to select a 
particular linear basis in $\mathcal H$. Actually, three of them turn 
out to be quite useful. The first one, the ``PBW-basis'', is given 
(up to ordering) by the set of monomials $X_{+}^a X_{-}^b K^c$.

The second one, that we shall call the ``elementary basis'' comes from 
the previous isomorphism with $M(3,\CC) \oplus \Mlo{2|1}$.
We call $E_{ij}$ the elementary matrices corresponding to the 
$M(3,\CC)$ block (they correspond to the $e_{ij}$ coefficients of
(\ref{structure_of_H_matrices})). As for the $\Mlo{2|1}$ block, we 
call $A_{ij},B_{ij},P_{ij},Q_{ij}$ the elementary matrices corresponding
to the $\alpha_{ij},\beta_{ij},\gamma_{ij},\delta_{ij}$ 
coefficients, respectively. Clearly, this set of elementary matrices
is also a basis of $\mathcal H$ and it is not too difficult (though 
it is cumbersome) to express each of its elements in terms of the 
PBW-basis.

The last useful basis, directly related to the elementary basis, 
is defined in Section~\ref{s:Killing_scalar_product_on_H}; it has the 
property of diagonalizing the ``hermitianized'' Killing form. 

%%%%%%%%%%%%%%%%%%%%%%%%%%%%%%%%%%%%%%%%%%%%%%%%%%%%%%%%%%%%%%%%%%%%%%%%%%
\newpage

\subsection{The Killing form on a quantum group}
\label{a:Killing_form}

\subsubsection{The adjoint representation of a quantum group} 

If $X \in H$, then the adjoint map $ad_{X}: H \mapsto H$ is defined by
$$
   ad_{X}(Y) \doteq X_{1} \, Y \, S(X_{2}) \ .
$$
Notice that this definition generalizes both the notion of adjoint 
representation for groups (where $\Delta g = g\otimes g$ and 
$S(g)=g^{-1}$, $g$ being a group element) and for Lie algebras (where 
$\Delta X = X \otimes \one + \one \otimes X$ and $S(X)=-X$, $X$ being 
a Lie algebra element).

\paragraph{The representation $ad$ is a left action.}
It is indeed easy to show that
$$
ad_{XY}(Z) = ad_{X}(ad_{Y}(Z)) \ .
$$

\noindent
Actually, it is also possible to define ``another'' adjoint 
representation by replacing the previous definition by 
$S(X_{1}) Y X_{2}\,$; this is not a left action but a right one (so 
it can be called the ``right''-adjoint action).

One could be tempted to consider the right action $S^{-1}(X_1)YX_2$ 
or the left action $ X_{1} Y S^{-1}(X_{2})$ but these actions are not 
compatible with the algebra structure (indeed acting on the unit with 
some element $X$ would not give $\epsilon(X) \, \one$).
Moreover it is not very useful to consider the left and right actions 
$X_2 Y S^{-1}(X_1)$ and $S^{-1}(X_2)YX_1$ since, although compatible 
with the algebra structure, they are essentially equivalent with the 
previously given definitions for the left and right adjoint actions.
In fact the antipode intertwines both maps.

In the sequel, we shall only use the first definition of the adjoint 
action, we should therefore remember that it is a left action.

\paragraph{The adjoint action is compatible with the algebra structure.}
One indeed shows that
$$
ad_{X}(YZ) = ad_{X_1}(Y) \, ad_{X_2}(Z) \ .
$$

\noindent
Notice that the two given properties allow one to compute easily the 
explicit expression for the adjoint representation once it is known 
on the generators.

\paragraph{Case of ${\mathcal H}$.} 
In this case, one obtains easily the adjoint action on the generators:

$$
\begin{tabular}{ll}
$ad_{K}(K) = K$ &
   $ad_{X_{-}}(K) = (1-q^{-2}) X_{-} K^{2}$ \cr
$ad_{K}(X_{-}) = q^{-2} X_{-}$ &
   $ad_{X_{-}}(X_{-}) = 0$ \cr
$ad_{K}(X_{+}) = q^{2} X_{+}$ &
   $ad_{X_{-}}(X_{+}) = \frac{\one-K^2}{q-q^{-1}}$ \cr
$ad_{X_{+}}(K) = (1-q^{2}) X_{+} K$ &
   $ad_{X_{+}}(X_{+}) = (1-q^{2})X_{+}^{2}$ \cr
$ad_{X_{+}}(X_{-}) = (1-q^{-2}) X_{+}X_{-} + \frac{K-K^{-1}}{q^3-q}$ & \cr
\end{tabular}
$$

%%%%%%%%%%%%%%%%%%%%%%%%%%%%%%%%%%%%%%%%%%%%%%%%%%%%%%%%%%%%%%%%%%%%%%%%%% 

\subsubsection{The quantum trace} 

If $H$ is a quasitriangular Hopf algebra, with an universal $R$-matrix 
$\mathcal R$, there exists in it a special element
$$
   u_o \doteq m[(S\otimes \id) \mathcal{R}_{21}] \ .
$$
Such an element is invertible and allows to write explicitly $S^2$ as
an inner automorphism (see \cite{Chari-Pressley} for a proof and a more 
general discussion):
$$
   S^2(h) = u_o hu_o^{-1} \ , \qquad \forall h \in H \ .
$$

On the other hand, given $\rho$ a representation of $H$ on a space $V$, 
the quantum trace is the map defined by the following chain of 
isomorphisms, all of them commuting with the $H$-action:
$$
   End(V) \rightarrow V \otimes V^\star \rightarrow 
                      V^{\star\star} \otimes V^\star \rightarrow
                      \CC \ .
$$
Remember that given a representation on $V$, one obtains naturally a 
representation on its dual space $V^\star$, by making use of the antipode 
($h \triangleright v^\star$ is such that 
$\langle h \triangleright v^\star \,, w \rangle = 
    \langle v^\star \,, S(h) \triangleright w \rangle 
    \ \forall w \in V$).
The non-canonical isomorphism $V \simeq V^{\star\star}$ given by
$v \rightarrow \rho(u_o) v$ is needed in order to make the chain commute 
with the action of the quantum group.
Therefore, the resulting expression for the quantum trace in terms 
of the ordinary operator trace on $V$ is
$$
   \tr_q(X) = \tr(\rho(u_o)X) \ , \qquad X\in End(V) \ .
$$
As $u_o$ has no reason to be group-like, this trace is in general not 
multiplicative on tensor products of representations of $H$, but 
can be made so if $H$ is a ribbon Hopf algebra. In this case there 
exists an invertible and central element $v \in H$ such that 
$v^2 = u_o S(u_o)$, $S(v)=v$, and 
$\Delta v = (\mathcal{R}_{21}\mathcal{R}_{12})^{-1} (v\otimes v)$.
Now $u_o$ may be replaced in $\tr_q$ by $u \doteq v^{-1} u_o$, which is 
group-like. It is still true that $S^2(h) = u h u^{-1}$, because $v$ is 
central.

In the case of $H=\mathcal{H}$ we find $u=K^{-1}$ (and $v=\one$).

%%%%%%%%%%%%%%%%%%%%%%%%%%%%%%%%%%%%%%%%%%%%%%%%%%%%%%%%%%%%%%%%%%%%%%%%%% 

\subsubsection{The Killing form}

Let $X$, $Y$ denote two matrices (with $\CC$-number entries!) 
representing elements $X$ and $Y$ of a Hopf algebra $H$ in some 
representation (we keep the same notation, here, for elements of the 
Hopf algebra and their matrix representatives). The Killing form 
in this representation is defined by
$$
   \left( X, Y \right)_u \doteq \tr_{q}(X Y) = \tr(u X Y) \ .
$$
The terminology ``Killing form'' usually refers to a particular 
bilinear form on a Lie algebra and its representations. 
Extension of this notion to the enveloping associative algebra is 
usually not considered. In the present case, we are therefore using a 
slightly generalized terminology (like in \cite{Schupp-Watts-Zumino}). 
Notice that in our examples the Hopf algebra $H$ is finite dimensional, 
so we can even discuss the structure of this Killing form in the regular 
representation.

%%%%%%%%%%%%%%%%%%%%%%%%%%%%%%%%%%%%%%%%%%%%%%%%%%%%%%%%%%%%%%%%%%%%%%%%%% 

\paragraph{Symmetry of the Killing form.}
As $S^{2}(X)=u X u^{-1}$, then
\begin{eqnarray*}
\left( X,S^{2}(Y) \right)_u &=& \left( X,u Y u^{-1}\right)_u = 
                               \tr_{q}(u X u Y u^{-1}) \cr
                            &=& \tr_{q}(X u Y) = \tr_{q}(u Y X) \ .
\end{eqnarray*}
Therefore
\begin{equation}
\left( Y, X \right)_u = \left( X,S^{2}(Y) \right)_u \ . 
\label{symmetry_of_Killing}
\end{equation}
This reduces to the usual symmetry when $S^{2}$ is the identity, 
which is in particular the case for a group. 

\paragraph{Invariance of the Killing form under the adjoint action.}
One can show that
\begin{equation}
\left( ad_{Z_1}(X), ad_{Z_2}(Y) \right)_u = 
    \left( X,Y \right) _{u} \, \epsilon (Z) \ .
\label{invariance_of_Killing}
\end{equation}
In the classical case of a group or a Lie algebra, this reduces to 
the usual invariance of the Killing form under the adjoint action. 

To prove this property, one needs the following lemma:
$$ 
\tr_{q}(u \, ad_{X}(Y)) = \tr(u Y) \, \epsilon(X) \ .
$$
Indeed,
\begin{eqnarray*}
\tr_{q}(u \, ad_{X}(Y)) &=& \tr_{q}(u X_{1} Y S(X_{2})) = 
                             (X_{1},Y S(X_{2}))_u =
                             (Y S(X_{2}), S^{2}(X_{1}))_u \\
                      {} &=& \tr(u Y S(X_{2}) S^{2}(X_{1})) \\
                      {} &=& \tr(u Y) \, \epsilon(X) \ .
\end{eqnarray*}
Therefore, the left hand side of (\ref{invariance_of_Killing}) reads 
\begin{eqnarray*}
   \tr_{q}(u \, ad_{Z_1}(X) \, ad_{Z_2}(Y)) &=& 
   \tr_{q}(u \, ad_{Z}(XY)) = \tr(u \, X Y) \, \epsilon(Z) \\
   &=& (X,Y)_u \, \epsilon(Z) \ .
\end{eqnarray*}

%%%%%%%%%%%%%%%%%%%%%%%%%%%%%%%%%%%%%%%%%%%%%%%%%%%%%%%%%%%%%%%%%%%%%%%%%% 
%%%%%%%%%%%%%%%%%%%%%%%%%%%%%%%%%%%%%%%%%%%%%%%%%%%%%%%%%%%%%%%%%%%%%%%%%% 
\newpage

\subsection{The ``double'' $\tilde\mathcal{H}$ of $\mathcal{H}$} 
\label{a:algebras_related_to_H}

We now take $q^{N}=1$ ($N$ odd), as before, but consider the finite 
dimensional quotient $\tilde\mathcal{H}$ of the quantum algebra 
$U_{q}(sl(2,\CC))$ by the Hopf ideal defined by $X_{\pm}^N=0$, 
$K^{2N}=\one$ (rather than $K^{N}=\one$). Notice that this ``double'' 
has nothing to do with what is called the ``quantum double'' of a 
Hopf algebra in the literature. 

In order to make use of all the results concerning $\mathcal H$, take 
$X_{\pm}$ and $K$ as the generators of $\mathcal H$, as before, and call 
$\tilde X_{\pm}$ and $\tilde K$ the generators of $\tilde\mathcal{H}$. 
Now set
\begin{eqnarray}
\tilde K &=& \sigma_{3} \otimes K = \diag(K, -K) \nonumber \\
\tilde X_{+} &=& \one \otimes X_{+} = \diag( X_+, X_+) 
   \label{double_of_H} \\
\tilde X_{-} &=& \sigma_{3} \otimes X_{-} = \diag(X_-, -X_-) \ , \nonumber
\end{eqnarray}
where $\sigma_{i}$ are the Pauli matrices. This provides an explicit 
realization of $\tilde\mathcal{H}$ in terms of ${\mathcal H}$. One 
sees immediately that 
$dim (\tilde\mathcal{H}) = 2\,dim(\mathcal{H}) = 2 N^{3}$ 
and obtains also for $N=3$ an explicit expression for the generators, 
in terms of Grassmann valued $12\times 12$ matrices, by using the 
expressions of $X_{\pm},K$ given in \cite{Coquereaux} or \cite{CoGaTr-e}. 
By construction, it is clear that $\mathcal{H}$ is a $\ZZ_{2}$ quotient 
of $\tilde\mathcal{H}$ ---notice that the group generated by powers of 
$\tilde K$ is no longer $\ZZ_{3}$, like before, but 
$\ZZ_{3}\times \ZZ_{2}$ and that ${\tilde K}^{3}$ is a non trivial 
central element. 

The representation theory of this algebra can then be obtained in 
a straightforward manner: projective indecomposable representations 
are still given by the columns of the corresponding isomorphic Grassmann 
valued matrix algebra; the ones appearing in the upper diagonal 
$6\times 6$ block of (\ref{double_of_H}) are the same $3_{irr}$, 
$6_{odd}$ and $6_{eve}$ considered in \ref{a:structure_of_H}; those 
appearing in the lower block will be denoted by $3^-_{irr}$, $6^-_{odd}$ 
and $6^-_{eve}$. More generally (for arbitrary $N$), we see that 
indecomposable representations of $\tilde\mathcal{H}$ are of two kinds: 
they can be labeled by $\omega = \pm 1$, those for which $\omega = 1$ 
are {\sl also} representations of $\mathcal{H}$, whereas those for which 
$\omega = -1$ only appear as representations of $\tilde\mathcal{H}$. 
These two kinds of representations can therefore be distinguished by the 
eigenvalue of the non trivial central element ${\tilde K}^{3}$. Remark 
that, when $\tilde \mathcal H$ is (faithfully) realized, as explained 
above, in terms of $12 \times 12$ matrices with Grassmann entries, the 
restrictions $K|_{1}$ and $K|_{2}$ of $\tilde K$ to the upper and lower 
blocks are such that  $\tilde{K}|_{1}^{\;3} = \one_{3\times 3}$, and 
$\tilde{K}|_{2}^{\;3} = -\one_{6\times 6}$.

It may be useful to recall that, when $q$ is an odd ($N$) root of unity, 
the center of ${U}_q(sl(2,\CC))$ is generated by the Casimir $C$, 
$X_{\pm}^{N}$ and $K^{\pm N}$. Call $c,x,y$ and $z^{\pm 1}$ the values of 
these central elements in irreducible representations. There are 
irreducible representations ``of classical type'' usually denoted by 
$Spin(j,\omega)$, where $j$ is a half-integer spin and $\omega = \pm 1$; 
in those representations $x = y = 0$ and $z = \omega^{N} = \pm 1$. There 
are also irreducible representations ``of non classical type'' which can 
be ``periodic'' ($xy \neq 0$) or semi-periodic ($xy = 0$ but either 
$x \neq 0$ or $y \neq 0$); such representations do not appear for finite 
dimensional Hopf algebras quotients such as $\mathcal H$ since since 
both $x$ and $y$ will then automatically vanish. Somehow, considering 
$\tilde \mathcal H$ instead of $\mathcal H$ has the interest of allowing 
one to recover also the irreducible representations of 
${U}_q(sl(2,\CC))$ with $\omega = -1$ as representations of a 
quasitriangular {\sl finite} dimensional Hopf algebra.

A general discussion concerning Hopf stars, twisted or not, and scalar 
products, can be done here along the same general lines as before. In 
particular, notice that when we chose one of the two possible twisted 
Hopf stars ($X_+^* = \pm X_-$, $X_-^* = \pm X_+$, $K*= K^{-1}$), the 
invariant scalar products associated with the family of corresponding 
star representations ($\omega = \pm 1$) of $\tilde \mathcal H$ 
simultaneously exhibit features that in the case of $\mathcal{H}$ were 
obtained separately for (twisted) stars of type $SU(2)$ or $SU(1,1)$. 
For example, we know (see 
Section~\ref{s:twisted_metrics_on_H_representations}) that the invariant 
scalar product on $3_{irr}$ (\ie $\omega = +1$) associated with the 
$SU(2)$ twisted Hopf star is of signature $(++-)$, and that the 
signature is $+++$ for the twisted star of type $SU(1,1)$. It happens 
that the conclusions are just to be reversed when we replace $3_{irr}$ 
by $3^-_{irr}$ $(\omega = -1)$. It may also be of interest to notice 
that invariant scalar products corresponding to irreducible 
representations $3_{irr}^-$ and $2_{eve}$ (for the twisted $SU(2)$ case), 
or $3_{irr}$ and $2_{eve}^-$ (for the twisted $SU(1,1)$ case) of this 
double $\tilde\mathcal{H}$ have a positive definite metric.

Regarding invariant scalar products on the left regular representation 
of $\tilde\mathcal{H}$ (as a module-algebra), we have a freedom of $54$ 
real parameters, for the same reasons as those given in 
Section~\ref{s:scalar_product_on_H}, but specific scalar products 
can be defined as in Section~\ref{s:scalar_products_on_left_rr}.

\paragraph{Remark: The simply connected form of $U_q(sl(2,\CC))$.} 
A standard construction at the level of the infinite dimensional 
universal quantum algebra (see for example \cite{Chari-Pressley, Klimyk})
$U_q(sl(2,\CC))$, consists in introducing a square root $k$ for $K$, 
so that $k^2=K$; it is also useful to define generators $I_\pm$ for which 
the coproduct is symmetrical, \ie 
$\Delta I_{\pm}=I_{\pm}\otimes k^{-1} + k \otimes I_{\pm}$. 
This infinite dimensional algebra generated by $\{k,I_\pm\}$ is often 
called the ``simply connected form'' of $U_q(sl(2,\CC))$ and denoted 
$\check{U}_q(sl(2,\CC))$ (warning: in the literature this object is 
sometimes called just $SL_q(2)$!). $U_q(sl(2,\CC))$ is a Hopf subalgebra 
of $\check{U}_q(sl(2,\CC))$; the explicit inclusion of the former in the 
latter can be obtained by taking $K = k^{2}$, $X_{+} = I_{+} k$ and 
$X_{-} =k^{-1} I_{-}$. Since $k^{3}$ is central, one could then be 
tempted to build a finite dimensional Hopf quotient of 
$\check{U}_q(sl(2,\CC))$ by factoring it by the ideal given by 
$I_{\pm}^{3} = 0$ and $k^{3}=1$. The point is that one does not get 
anything essentially new by doing so: the obtained quotient is isomorphic 
with $\mathcal H$ itself. Indeed, let us set $K \doteq k^{2}$ at the level 
of this quotient, then $K^{2}=k^{4}=k$ and $K^{3}=k^{6}=\one$. Hence the 
relation between $k$ and $K$ can be inverted. Moreover, one can check 
explicitly (thanks to the previously given change of variables between 
$X_{\pm}$ and $I_{\pm}$) that all the algebra and coalgebra relations of 
this finite dimensional quotient of $\check{U}_q(sl(2,\CC))$ coincide 
exactly with those given for $\mathcal H$ itself. 

%%%%%%%%%%%%%%%%%%%%%%%%%%%%%%%%%%%%%%%%%%%%%%%%%%%%%%%%%%%%%%%%%%%%%%%%%% 
%%%%%%%%%%%%%%%%%%%%%%%%%%%%%%%%%%%%%%%%%%%%%%%%%%%%%%%%%%%%%%%%%%%%%%%%%% 
%%%%%%%%%%%%%%%%%%%%%%%%%%%%%%%%%%%%%%%%%%%%%%%%%%%%%%%%%%%%%%%%%%%%%%%%%% 

%%%%%%%%%%%%%%%%%%%%%%%%%%%%%%%%%%%%%%%%%%%%%%%%%%%%%%%%%%%%%%%%%%%%%%%%%% 

\end{document}